\documentclass[reprint,superscriptaddress,amsmath,amssymb,aps,prx]{revtex4-2}

\usepackage{graphicx}
\usepackage{dcolumn}
\usepackage{bm}
\usepackage{hyperref}
\usepackage{booktabs}
\usepackage{multirow}
\usepackage{float}
\usepackage{longtable}
\DeclareUnicodeCharacter{00B2}{\textsuperscript{2}}
\DeclareUnicodeCharacter{03B8}{\ensuremath{\theta}}

\begin{document}

\title{Attention Is Not All You Need for Diffraction}

\author{Elizabeth J. Baggett}
\affiliation{Department of Physics, Boston College, Chestnut Hill, MA 02467, USA}
\affiliation{NIST Center for Neutron Research, National Institute of Standards and Technology, Gaithersburg, MD 20899, USA}

\author{Edward G. Friedman}
\affiliation{Computer Science Department, School of Computer Science, Carnegie Mellon University, Pittsburgh, PA 15213, USA}
\affiliation{NIST Center for Neutron Research, National Institute of Standards and Technology, Gaithersburg, MD 20899, USA}

\author{Abhishek Shetty}
\author{Derrick Chan-Sew}
\author{Vanellsa Acha}
\author{Harshita Dwarcherla}
\affiliation{School of Information, University of California, Berkeley, CA 94720, USA}

\author{Paul Kienzle}
\affiliation{NIST Center for Neutron Research, National Institute of Standards and Technology, Gaithersburg, MD 20899, USA}

\author{William Ratcliff}
\email{william.ratcliff@nist.gov}
\affiliation{NIST Center for Neutron Research, National Institute of Standards and Technology, Gaithersburg, MD 20899, USA}
\affiliation{Department of Materials Science and Engineering, University of Maryland, College Park, MD 20742, USA}
\affiliation{Department of Physics, University of Maryland, College Park, MD 20742, USA}

\date{\today}

\begin{abstract}
Determining crystal symmetry from powder X-ray diffraction is a central problem in materials characterization, yet multiple space groups can produce indistinguishable patterns posing a significant challenge for automated classification. We show that attention-based architectures, while superior to convolutional networks for this task, are insufficient on their own: reliable symmetry extraction requires encoding crystallographic knowledge into both the network architecture and the training curriculum. We introduce a physics-informed transformer that classifies powder patterns into 99 extinction groups, the most specific symmetry classification accessible from diffraction data alone, using an explicit $\sin^2\!\theta$ coordinate channel, physics-aware positional encoding, and a structured multi-task decoder that separates geometric rule learning from holistic pattern recognition. A three-stage curriculum of balanced synthetic pretraining, realistic fine-tuning with explicit preferred-orientation modeling, and Bayesian prior injection proves essential for bridging the synthetic-to-real domain gap, while post-hoc temperature scaling---rather than additional training---is the key remaining ingredient for robust real-data transfer. By mapping predictions onto the directed acyclic graph of maximal \textit{translationengleiche} subgroups, we show that the calibrated model's errors are not random but physically structured: they remain local on the subgroup hierarchy and flow predominantly toward lower-symmetry descendants, consistent with the physical erasure of systematic-absence cues by real-world noise. We further identify a ``catastrophic paradox'' in which classical Rietveld fit quality does not cleanly predict neural classification difficulty, because lower label-space entropy partly offsets worse profile fits. These results establish that physics-informed target design, curriculum, and calibrated inference matter as much as model capacity for scientific machine learning on diffraction data.
\end{abstract}

\maketitle

\section{Introduction}

Powder X-ray diffraction (PXRD) is one of the most widely used probes for determining crystal structure, yet extracting symmetry information from a powder pattern remains a notoriously ill-posed inverse problem. Segal \textit{et al.}\ have shown that the PXRD-to-structure loss landscape is highly non-convex: commonly used similarity metrics produce rugged optimization surfaces, and gradient-based refinement can converge to incorrect local minima even from moderately distorted initial states~\cite{segal2023pxrd}. Constraining the search to the correct crystal family helps, but does not eliminate the problem. Traditionally, symmetry determination requires manual analysis by an expert crystallographer---a process poorly suited to the massive data volumes produced by modern high-throughput experiments and autonomous beamlines.

Machine learning offers a path toward automating this analysis. Convolutional neural networks (CNNs) were first applied to space-group classification from powder patterns by Park \textit{et al.}~\cite{park2017classification}, and subsequent work by Lolla \textit{et al.}~\cite{Lolla2022SemiSupervisedCrystal} and Schopmans \textit{et al.}~\cite{schopmans_neural_2023} achieved strong performance using ResNet architectures trained on synthetic data. More recently, transformer architectures~\cite{vaswani2023attentionneed} have been applied to diffraction: Chen \textit{et al.}~\cite{D3DD00198A} used a Vision Transformer (ViT)~\cite{dosovitskiy2021imageworth16x16words} to classify metal--organic frameworks from PXRD patterns, and Simonnet \textit{et al.}~\cite{10887635} reported gains over CNNs for mineral identification from synthetic single-phase data.

However, these prior studies share two limitations. First, they frame the task as space-group classification, even though diffraction cannot distinguish space groups that share identical reflection conditions. Second, they treat the diffraction pattern as a generic one-dimensional signal, applying standard image-classification architectures without encoding the physics of diffraction geometry. In this work, we address both limitations.

We find that reliable symmetry extraction requires physics at every level of the pipeline, not just the model architecture. Reframing the classification target from 230 space groups to 99 extinction groups---the information-theoretically identifiable equivalence classes under powder diffraction---more than doubles Top-1 accuracy on held-out synthetic data in controlled comparisons. A single post-hoc calibration parameter, applied without retraining, triples Top-1 accuracy on degraded real minerals by decoupling the model's learned evidence from the geological prior absorbed during fine-tuning. And when we map the calibrated model's residual errors onto the crystallographic subgroup hierarchy, we find they are not random: the model systematically falls to nearby lower-symmetry descendants, behaving like a conservative crystallographer under uncertainty. These results suggest that the $\sim$10\% Top-1 ceiling on highly degraded real mixtures reflects a fundamental information-theoretic constraint of 1D powder diffraction rather than a limitation of the model.

In this work, we argue that attention mechanisms, while necessary, are not sufficient for reliable symmetry extraction from powder data. We make three contributions:
\begin{enumerate}
\item We demonstrate that \textbf{extinction groups}---the 99 equivalence classes of space groups sharing identical systematic absences---are the information-theoretically correct classification targets for powder diffraction, and that reframing the task at this level substantially improves classification accuracy in synthetic benchmarks. A matched 230-space-group control confirms this point: even after collapsing its predictions post-hoc into extinction-group space, the space-group route remains well below direct extinction-group training on the same held-out regime.
\item We introduce a \textbf{physics-informed transformer architecture} that incorporates an explicit $\sin^2\!\theta$ coordinate channel, physics-aware positional encoding, and a structured multi-task decoder separating crystallographic rule learning from holistic pattern recognition.
\item We show that a \textbf{three-stage training curriculum}---balanced synthetic pretraining, RRUFF-conditioned realistic fine-tuning, and Bayesian prior injection at inference---is essential for bridging the synthetic-to-real domain gap. We further demonstrate that post-hoc temperature scaling (here, temperature is a mathematical scalar controlling the sharpness of the predicted probability distribution, not a physical quantity) resolves target-domain overconfidence, producing physically interpretable error structure on real-world mixtures.
\end{enumerate}

We evaluate all models on a new algorithmically curated benchmark of 473 real RRUFF mineral patterns and report a detailed analysis of failure modes, including a ``catastrophic paradox'' in which classical Rietveld fit quality does not cleanly predict neural classification difficulty.

\section{The Case for Extinction Groups}
\label{sec:extinctiongroups}

An extinction group is the set of space groups that produce identical systematic absences in reciprocal space. Symmetry elements such as lattice centering, glide planes, and screw axes determine which Miller indices $(hkl)$ can give rise to diffraction peaks. Because different space groups can share identical reflection conditions, they are experimentally indistinguishable by standard powder diffraction. For example, Friedel's law further eliminates the distinction between centrosymmetric and non-centrosymmetric structures when anomalous scattering is negligible. These overlaps reduce the 230 crystallographic space groups to 99 unique extinction groups. (We merge hexagonal $P$--$c$-- and trigonal $P$----$c$ because they produce identical systematic absences.)

The collapse from 230 space groups to 99 extinction groups has a direct consequence for machine learning: any model trained to predict space groups is penalized by the loss function for failing to distinguish physically indistinguishable patterns. Table~\ref{tab:spgvsexg} illustrates the effect: switching from space-group to extinction-group classification with the same ResNet-18 architecture substantially improves Top-1 accuracy (37\% to 80\%). This first comparison was deliberately illustrative, because it was confounded by both class count and dataset size. An additional, cleaner control matched 2.0M uniform corpus (see Supplemental Material for full details). Once both models are scored in extinction-group space, the post-hoc SG$\rightarrow$EG collapse reaches 8.61\% Top-1 versus 19.32\% for direct EG training---a factor-of-two advantage that confirms the target-design effect is not merely a class-count artifact.

This mirrors standard crystallographic practice, in which an experimentalist first determines a pool of candidate space groups from extinction conditions, then narrows that pool using non-diffraction constraints such as polarity, centrosymmetry, and formula-unit compatibility with the unit-cell volume.

\begin{table}[h]
\caption{ResNet-18 classification accuracy on balanced synthetic reflection data. The two runs differ in class count and dataset size (2.3M across 230 space groups vs.\ 990k across 99 extinction groups), so this table should be read as an illustrative rather than controlled comparison. The matched 2.0M SG$\rightarrow$EG control discussed in the text provides that cleaner comparison and still favors direct extinction-group training.}
\begin{ruledtabular}
\begin{tabular}{lccc}
Target & Top-1 & Top-3 & Top-5 \\
\colrule
Space Groups (230 classes)     & 37\% & 74\% & 88\% \\
Extinction Groups (99 classes) & 80\% & 95\% & 97\% \\
\end{tabular}
\end{ruledtabular}
\label{tab:spgvsexg}
\end{table}

\section{Data}
\label{sec:data}

\subsection{The Distribution Problem}

\begin{figure}[h]
\centering
\includegraphics[width=0.45\textwidth]{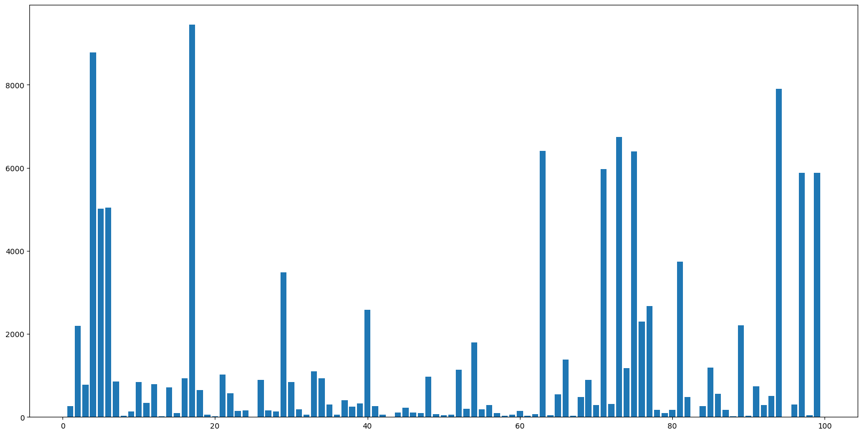}
\caption{Extinction-group distribution in the Inorganic Crystal Structure Database (ICSD), showing severe geological class imbalance.}
\label{fig:icsd_distribution}
\end{figure}

Well-populated crystal structure databases such as the ICSD exhibit steep label imbalance (Fig.~\ref{fig:icsd_distribution}). Table~\ref{tab:distribution} illustrates the consequences: when a CNN is evaluated on the RRUFF test set, models trained on biased distributions outperform the frequency baseline at Top-1, indicating genuine feature learning, but by $k\approx 5$ their rankings increasingly recapitulate the training prior rather than discriminating among candidates on the basis of the pattern of systematic absences.

\begin{table}[h]
\caption{Top-$k$ accuracy on the RRUFF test set when sampling training labels from different distributions. Above $k \approx 5$, models trained on biased distributions are largely recapitulating the prior.}
\begin{ruledtabular}
\begin{tabular}{cccc}
$k$ & RRUFF & ICSD  & Uniform \\
\colrule
1  & 12.3\% & 9.7\%  & 1.0\%  \\
5  & 47.4\% & 27.2\% & 5.1\%  \\
10 & 67.0\% & 53.7\% & 10.1\% \\
20 & 83.7\% & 76.4\% & 20.2\%   \\
\end{tabular}
\end{ruledtabular}
\label{tab:distribution}
\end{table}

Cross-evaluation confirms that balanced training produces more robust models: when a biased model is tested on a balanced test set, Top-1 drops to 3\%, whereas a balanced model tested on the biased set retains 8\% Top-1 (Table~S1, Supplemental Material).

\subsection{Synthetic Data Generation}

We employ two complementary generation strategies, both producing extinction-group-balanced datasets.

\textbf{Reflection-Based Generation.} For each target space group we generate random lattice parameters consistent with its metric constraints and use the Computational Crystallography Toolbox (\texttt{cctbx})~\cite{ccbtx} to compute the allowed reflections. Each reflection is represented by a Gaussian peak at its $2\theta$ position with width given by the instrument resolution function.

\textbf{Crystal-Structure-Based Generation.} We use \texttt{PyXtal}~\cite{pyxtal} to build synthetic crystal structures, verified with \texttt{ASE}~\cite{ase-paper} to ensure the intended symmetry is preserved. Powder patterns are then simulated with \texttt{PyCrysFML}~\cite{pycrysfml2023}, which computes full structure-factor intensities together with Lorentz-type intensity weighting and instrument-specific resolution broadening, producing more realistic intensity envelopes at the cost of higher computational expense.

Full details of the database architecture, parallel generation pipeline, and \texttt{HDF5} storage scheme are given in the Supplemental Material.

\section{Architecture}
\label{sec:architecture}

Powder diffraction is not a translation-invariant signal: absolute peak position matters because it encodes reciprocal-space geometry, and small lattice changes shift the entire pattern nonlinearly through Bragg's law. The position-dependence of the signal is the core reason CNNs plateau early on this task, while attention-based models perform better on both synthetic scaling studies and downstream real-data transfer. Beyond any single peak's absolute location, what carries diagnostic weight is the relational structure among peaks -- their pairwise distances, intensity ratios, and co-occurrence patterns -- dependencies that attention mechanisms are explicitly designed to capture, but that convolutional filters, constrained by local receptive fields, cannot model without significant depth. Our model therefore starts from a transformer backbone, extending the standard encoder with three physics-motivated modifications.

\subsection{Coordinate Channel}

Instead of feeding the transformer a single intensity channel, we provide a two-channel input: the intensity profile and an explicit $\sin^2\!\theta$ coordinate grid. This acts as a physical ruler that encodes the metric tensor relationship between peak position and $d$-spacing, sparing the model from having to infer diffraction geometry from positional encoding alone.

\subsection{Physics-Aware Positional Encoding}

For the ViT models used in the real-data curriculum, we retain a learned absolute positional embedding over patch tokens and add a physics-derived positional term aligned to diffraction geometry. Concretely, each patch is assigned its mean $2\theta$ coordinate, transformed to $\sin^2\!\theta$, scaled, and passed through a small MLP to produce a patch-level embedding that is added to the learned positional embedding. This injects reciprocal-space structure directly into the token positions while preserving the flexibility of a learned absolute embedding. In the trained real-data model, this learned physics term is nearly one-dimensional and monotone in the input $\sin^2\!\theta$ coordinate, indicating that it functions primarily as a learned reciprocal-space ruler rather than as an arbitrary positional code (see Supplemental Material).

This mechanism is distinct from the coordinate channel above. The coordinate channel exposes pointwise $\sin^2\!\theta$ values as an additional input feature before patchification, allowing early layers to reason jointly over intensity and physical location at the raw-trace level. The physics-aware positional encoding instead acts at the patch-token level, biasing self-attention with diffraction-aware token positions. In practice, the two components play complementary roles: one is an input-side physical ruler, while the other is a token-level positional prior.

\subsection{Dual-Head Decoder}
\label{sec:dualhead}

The shared transformer backbone feeds into two decoders that are trained jointly.

\textbf{Split Head (Crystallographic Rule Decoder).} This head predicts a structured set of crystallographic bits: crystal system, lattice centering, and glide-plane/screw-axis features, decoded through a deterministic lookup into extinction groups. The split head outputs a 37-bit target vector; crystal system and centering are trained with cross-entropy losses, while the sparse operator bits use binary cross-entropy with positive-class weighting (\texttt{pos\_weight}). Not all 37-bit combinations correspond to valid extinction groups: the lookup table maps valid combinations to one or more extinction groups, while invalid bit patterns have no crystallographic interpretation. In practice, this loss acts as a structural regularizer: it forces the backbone to attend to weak present/absent peaks---the same features a crystallographer would examine---rather than relying solely on dominant fingerprint peaks.

\textbf{Auxiliary Head (Direct Extinction-Group Classifier).} This head predicts the 99 extinction groups directly via softmax. It learns the continuous joint distribution and is more robust to real-world noise than strict Boolean rule decoding.

\textbf{Fusion Decoder.} At inference time, we fuse the predictions:
\begin{equation}
\mathbf{p}_{\mathrm{fused}} = \alpha \, \mathbf{p}_{\mathrm{split}} + (1 - \alpha) \, \mathbf{p}_{\mathrm{aux}}
\label{eq:fusion}
\end{equation}
When rule evidence is weak, the split-head logits remain near zero and the induced extinction-group distribution becomes diffuse after lookup. Fusion therefore acts as a practical uncertainty-balancing mechanism: the auxiliary head dominates when the rule path is indecisive, while the split path sharpens predictions when physically distinctive absence cues are present.

\subsection{Model Specifications}

\textbf{Regular Transformer (RT).} $d_{\mathrm{model}} = 256$, 4 attention heads, RoPE~\cite{su2023roformerenhancedtransformerrotary}, adaptive average pooling, 2.73M parameters. AdamW optimizer~\cite{loshchilov2019decoupledweightdecayregularization} with step scheduler.

\textbf{Vision Transformer (ViT).} Patch size 25, learnable CLS token, 8 heads, $d_{\mathrm{model}} = 256$, 9.52M parameters. Adam optimizer~\cite{kingma2017adammethodstochasticoptimization}, fixed learning rate.

\section{Training Curriculum}
\label{sec:curriculum}

The central lesson of this work is that architecture and data design must be co-optimized. The main curriculum uses 1.38M uniform stage-1 samples followed by 2.35M corrected samples conditioned on RRUFF, an experimental mineral database, in stage 2. Late ablations then extend stage~2 with exact preferred orientation and mixed standard+preferred-orientation curricula (see Supplemental Material for full lineage).

\subsection{Phase 1: Uniform Synthetic Pretraining}

To prevent the model from collapsing toward common low-symmetry classes, we pretrain on a large synthetic dataset balanced uniformly across all 99 extinction groups. Because empirical mineral distributions are strongly skewed toward lower-symmetry classes, training directly on geological frequencies risks learning the prior before learning the weak diffraction signatures of rare centering, glide, and screw rules. Uniform pretraining forces the model to represent all extinction groups evenly, building a symmetry-balanced geometric engine.

\subsection{Phase 2: RRUFF-Style Synthetic Fine-Tuning}

Purely uniform synthetic data is too clean and chemically random. The second phase fine-tunes on synthetic patterns that incorporate realistic backgrounds, multiplicative and additive noise, peak broadening, and impurity phases. Crucially, main-phase stoichiometries are constructed from Wyckoff-compatible multiplicities, and structures are accepted only when the realized extinction group detected by \texttt{spglib} matches the intended label. This stage was the single largest driver of real-data improvement in our ablations.

\subsection{Phase 3: Bayesian Inference and Calibration}

Because Phase~1 deliberately removes geological priors, we restore them at inference time using an Empirical Bayes prior---concretely, the log-frequency of each extinction group in the training corpus, added to the auxiliary logits before softmax---and no evaluation labels are used (see Supplemental Material for provenance). However, the stage-2 model already absorbs part of the geological prior through fine-tuning on millions of RRUFF-conditioned patterns. Adding the external prior without calibration therefore effectively double-counts this bias. To resolve this, we apply post-hoc temperature scaling~\cite{guo2017calibration}---dividing the auxiliary logits by $T$ before adding the external log-prior---which proved critical for real-benchmark performance.

\begin{figure}[h]
\centering
\includegraphics[width=0.45\textwidth]{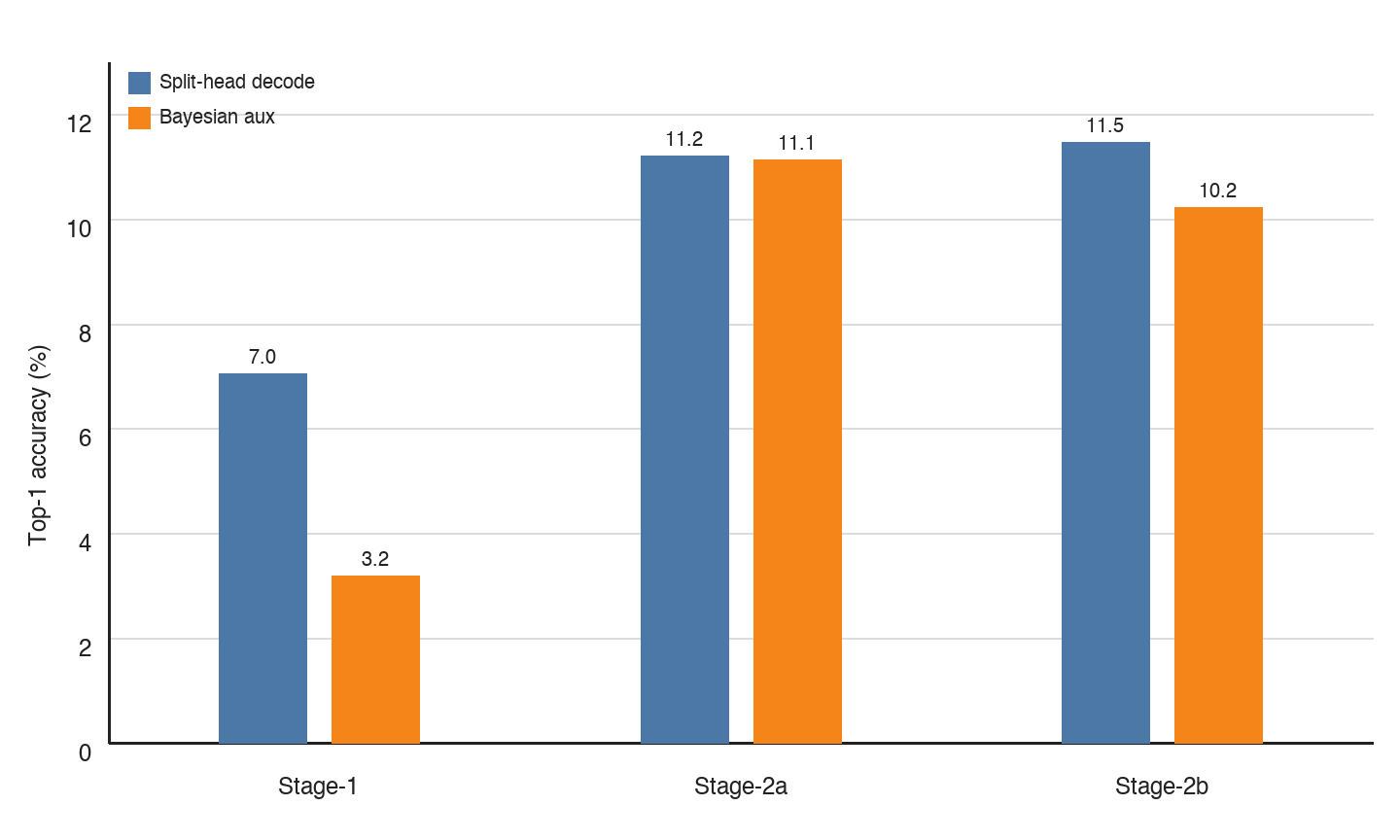}
\caption{Effect of curriculum stages on real RRUFF holdout Top-1 accuracy. RRUFF domain fine-tuning (Phase~2) is the dominant factor in bridging the sim-to-real gap.}
\label{fig:curriculum}
\end{figure}

Figure~\ref{fig:curriculum} shows the impact of each curriculum stage on a real RRUFF holdout.

\section{Results}
\label{sec:results}

We report results on three evaluation tiers: (i)~balanced synthetic data for architecture comparison and scaling, (ii)~the broader RRUFF-473 real benchmark for decoder comparison, and (iii)~the more challenging RRUFF-325 real benchmark for calibration and topological analysis.

\subsection{Scaling Studies}

Table~\ref{tab:scaling} presents the full scaling comparison on balanced reflection-based data. Despite having 3.5$\times$ fewer parameters, the regular transformer consistently outperforms the ViT on idealized synthetic benchmarks, suggesting that patch tokenization is a liability when the input consists of perfectly sharp peak profiles. By contrast, the ResNet-18 baseline plateaus at approximately 80\% Top-1 regardless of training set size (Fig.~S1, Supplemental Material).

However, the real-data system is built on the physics-informed ViT backbone. We interpret this as a transfer/generalization tradeoff: the ViT's patch-based representation is coarser, but that coarseness accommodates real-world peak broadening, background, and sample displacement. The ViT also provides a natural interface for the coordinate channel and physics-aware positional encoding, and its CLS token yields a global representation whose attention distribution can be visualized directly. Ablation studies (Table~S3, Supplemental Material) confirm that the ViT is less sensitive to distribution shifts in $2\theta$ range and generation method. A separate positional ablation on real data confirms that the physics-aware PE carries the stronger positional prior; removing it collapses strict split-head validity to zero on both RRUFF benchmarks (Supplemental Material). Supplemental analysis further shows that the learned physics-aware positional term behaves as an almost one-dimensional $Q^2$-like reciprocal-space ruler, supporting the interpretation that its gain comes from imposing the correct diffraction geometry rather than from adding arbitrary representational complexity.

\begin{table}[h]
\caption{Top-$k$ accuracy (\%) on balanced synthetic reflection data.}
\begin{ruledtabular}
\begin{tabular}{lcccccc}
Size & \multicolumn{3}{c}{Regular Transformer} & \multicolumn{3}{c}{Vision Transformer} \\
     & Top-1 & Top-3 & Top-5 & Top-1 & Top-3 & Top-5 \\
\colrule
99k  & 57.5 & 78.8 & 86.8 & 21.7 & 38.1 & 46.8 \\
396K & 80.4 & 94.2 & 97.1 & 30.9 & 49.4 & 59.6 \\
990K & 89.8 & 97.9 & 99.0 & 42.5 & 65.3 & 76.3 \\
2.17M& 93.7 & 99.2 & 99.7 & 60.9 & 85.1 & 91.9 \\
5.14M& 94.4 & 99.5 & 99.9 & 71.8 & 91.7 & 96.0 \\
\end{tabular}
\end{ruledtabular}
\label{tab:scaling}
\end{table}

\subsection{Real-Data Evaluation: Two Benchmarks}
\label{sec:realdata}

We use two complementary real-data benchmarks. The first is \textbf{RRUFF-473}, a 473-pattern benchmark built from upstream RRUFF scans~\cite{rruff2019} using algorithmic family-consistency filters and nuisance-fit stratification (172 recoverable, 153 usable, 74 poor, 74 catastrophic). This benchmark is algorithmically curated, not hand-cleaned, making the evaluation harder but more reproducible than sanitized sets used in prior work~\cite{schopmans_neural_2023}. The second is \textbf{RRUFF-325}, a deterministic downstream subset from the same curation pipeline that retains only the usable and recoverable nuisance-fit strata. We use this stricter slice for calibration and topological analysis so that the most extreme nuisance regimes do not dominate those controlled measurements.

\subsection{Decoder Comparison (RRUFF-473)}

\begin{figure}[h]
\centering
\includegraphics[width=0.45\textwidth]{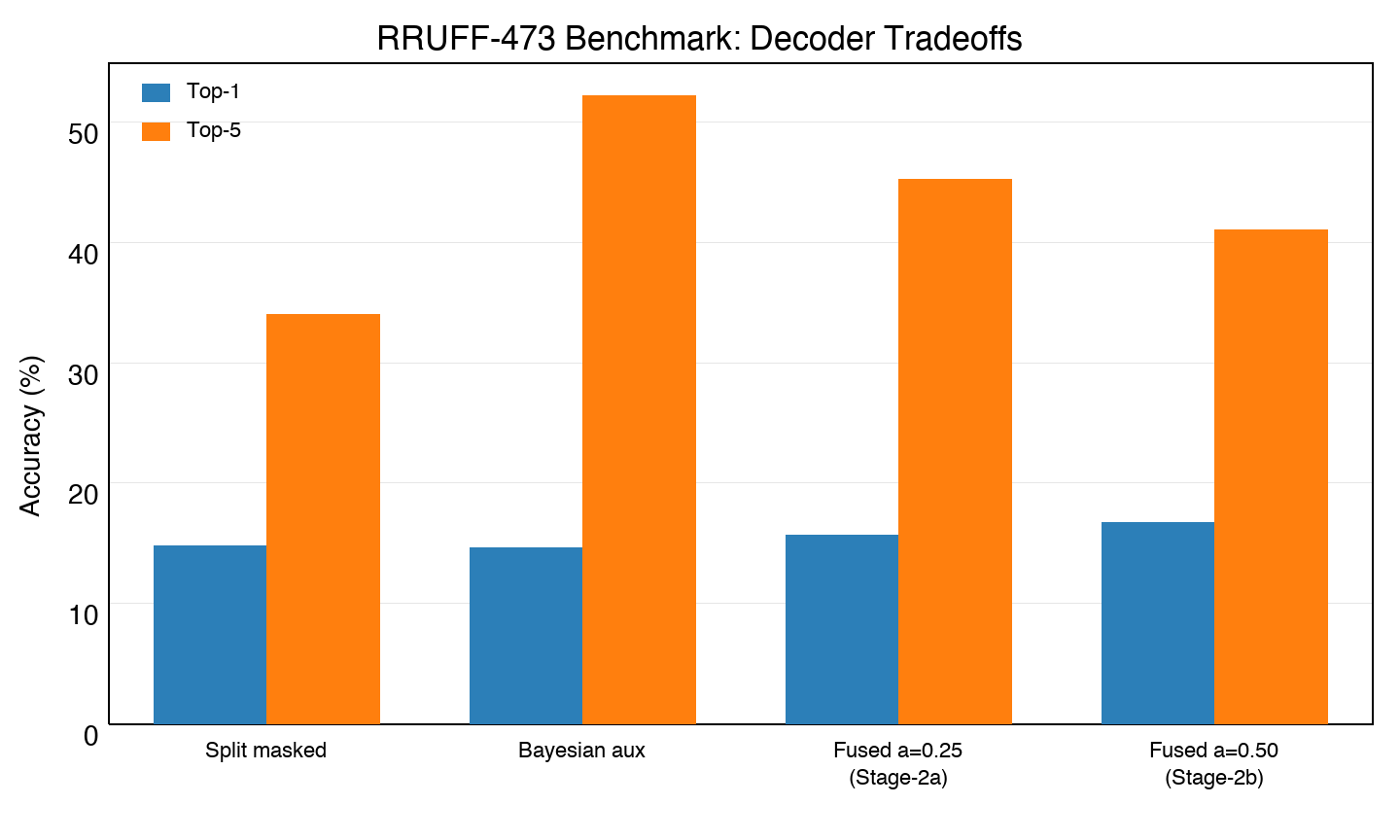}
\caption{Decoder comparison on the RRUFF-473 benchmark.}
\label{fig:decoder_tradeoffs}
\end{figure}

Table~\ref{tab:decoders} and Fig.~\ref{fig:decoder_tradeoffs} summarize the decoder comparison. The system has two distinct operating points. \textit{Best Top-1}: \textbf{Stage-2b} (larger uniform pretraining plus fine-tuning) with the fused decoder ($\alpha = 0.50$) at 16.70\%, benefiting from larger uniform pretraining that increased geometric rigidity in the split path. \textit{Best Top-5}: \textbf{Stage-2a} (the earlier fine-tuned checkpoint) with the Bayesian auxiliary head at 52.22\%, retaining the softer joint ranking that the larger uniform stage partially traded away. Fusion improves Top-1 for both checkpoints, confirming that the split and auxiliary paths provide complementary information.

\begin{table}[h]
\caption{Top-$k$ accuracy (\%) on the full RRUFF-473 benchmark.}
\begin{ruledtabular}
\begin{tabular}{llccc}
Model & Decoder & Top-1 & Top-3 & Top-5 \\
\colrule
\textbf{Stage-2a} & Split        & 14.80 & 28.96 & 34.04 \\
                   & Bayesian aux & 14.59 & 36.15 & 52.22 \\
                   & Fused ($\alpha\!=\!0.25$) & 15.64 & 34.88 & 45.24 \\
\colrule
\textbf{Stage-2b} & Split        & 15.01 & 27.91 & 32.77 \\
                   & Bayesian aux & 14.80 & 33.19 & 50.95 \\
                   & Fused ($\alpha\!=\!0.50$) & \textbf{16.70} & 32.98 & 41.01 \\
\end{tabular}
\end{ruledtabular}
\label{tab:decoders}
\end{table}

\subsection{Why the Split Head Is Still Valuable}

It may seem that the auxiliary head's dominance makes the split head redundant. This is incorrect, for three reasons. First, the split head acts as a structural regularizer for the shared backbone, preventing lazy shortcuts that overemphasize tallest peaks while ignoring weak symmetry-defining features. Second, the split and auxiliary heads fail in different ways, which is precisely why fusion improves Top-1. Third, the split head provides a geometry-driven backup when geological priors are misleading. The strongest evidence is the fusion result itself: if the split head contributed no information beyond what the auxiliary head already captured, fusion could not improve Top-1.

\subsection{Calibration on the Harsh 325-Pattern Benchmark}

\begin{figure}[h]
\centering
\includegraphics[width=0.45\textwidth]{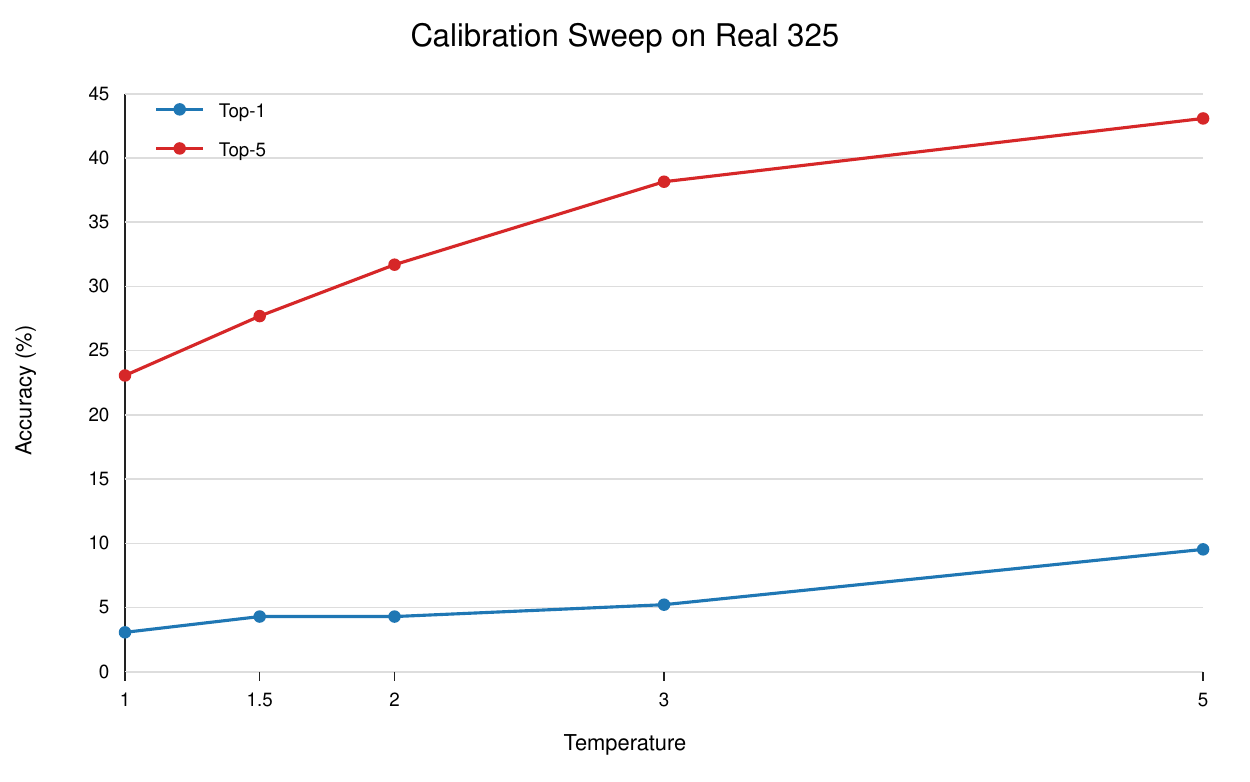}
\caption{Post-hoc calibration restores ranking accuracy on RRUFF-325. Results shown for the final Stage-2c checkpoint. Softening the auxiliary logits ($T=5$) before adding the geological prior decouples structural evidence from geological overconfidence.}
\label{fig:calibration_sweep}
\end{figure}

On RRUFF-325, the Stage-2c model's raw auxiliary logits are strongly overconfident: the uncalibrated auxiliary path reaches only 2.15\% Top-1 and 14.46\% Top-5 on the 325-pattern benchmark. Applying temperature scaling ($T=5$) to the auxiliary logits before Bayesian fusion changed the picture completely: the same checkpoint reached 9.54\% Top-1, 27.38\% Top-3, and 43.08\% Top-5 without changing any model weights (Fig.~\ref{fig:calibration_sweep}). Standard calibration diagnostics confirm this improvement: the expected calibration error (ECE) falls from 0.457 to 0.049, the negative log-likelihood (NLL) from 6.58 to 3.70, and the multiclass Brier score~\cite{brier1950verification} from 1.283 to 0.958.

This result materially changes the interpretation of stage~2. The fine-tuned model did learn useful real-domain structure, but its logits became saturated by the target-domain class imbalance. The dominant failure mode was not loss of diffraction physics; it was overconfident coupling of structural evidence to the geological prior. Temperature scaling restored usable uncertainty, allowing the external prior to act as intended rather than reinforcing the model's internal bias.

\subsection{Preferred Orientation and Mixed-Curriculum Ablations}

The stage-2c calibration result above established that post-hoc temperature scaling is essential, but the topological analysis (Section~\ref{sec:topology}) also exposed a physically interpretable failure mode: the calibrated model remained strongly biased toward lower-symmetry descendant predictions when experimental nuisance effects erased weak systematic-absence cues. To test whether subtractive experimental noise was underrepresented in the simulator, we added exact March--Dollase preferred orientation (PO), which models systematic reflection suppression due to non-random crystallite orientation, to the stage-2 realism curriculum while leaving stage~1 unchanged.

The pure-PO ablation produced the best exact ranking result on the harsh RRUFF-325 benchmark: after one epoch of PO-conditioned fine-tuning, the calibrated Bayesian auxiliary decoder reached 13.54\% Top-1 and 40.92\% Top-5 on RRUFF-325, and 13.53\% Top-1 and 49.89\% Top-5 on RRUFF-473. This confirms that explicit modeling of texture-suppressed reflections improves exact ranking on real diffraction patterns. However, it also narrowed the broader ranking distribution: on RRUFF-325, Top-5 remained below the earlier non-PO Stage-2c value of 43.08\%, and strict split-head validity collapsed to below 1\%, indicating that a pure texture-heavy curriculum destabilizes exact Boolean rule extraction.

We therefore trained a larger mixed stage-2 curriculum combining approximately 2.346M standard RRUFF-conditioned samples with 500k PO samples in a single one-epoch continuation from the same uniform base checkpoint. This large mixed model is the best balanced final model. On RRUFF-325 it achieved calibrated Top-1 / Top-5 = 10.46\% / 43.69\%, slightly exceeding the earlier Stage-2c Top-5 while still improving Top-1 over the non-PO baseline. On RRUFF-473 it achieved 9.94\% / 50.74\%. Most strikingly, split-head validity recovered from 0.92\% in the pure-PO run to 47.38\% on RRUFF-325 and 49.26\% on RRUFF-473, indicating that the mixed curriculum restores symbolic self-consistency while retaining the benefits of explicit PO exposure.

\begin{table*}[t]
\caption{Real-data performance and topological behavior on RRUFF-325 under calibrated Bayesian auxiliary decoding. All entries are measured on the same 325 patterns; ``Split valid'' reports the fraction for which the split head decodes to exactly one legal extinction-group template. Topological metrics (Desc./Anc./Branch, $\leq$2 hops, mean DAG distance) are defined in Section~\ref{sec:topology}; branch jumps denote errors that cross between distinct crystal-system families in the DAG rather than moving within a single family's hierarchy.}
\begin{ruledtabular}
\begin{tabular}{lccccc}
Model curriculum & Top-1 / Top-5 & Split valid & Desc./Anc./Branch & $\leq$2 hops & Mean DAG dist. \\
\colrule
Stage-2c (no PO)      & 9.54 / 43.08  & $\sim$0.6\% & 191 / 27 / 76 & 38.4\%  & 2.72 \\
PO-only, 1 epoch      & \textbf{13.54} / 40.92 & 0.92\% & 157 / 28 / 96 & \textbf{53.0\%} & 2.51 \\
Mixed 200k pilot      & 13.23 / 40.92 & 1.54\% & 153 / 30 / 99 & 49.65\% & 2.59 \\
Final large mixed run & 10.46 / \textbf{43.69} & \textbf{47.38\%} & 165 / 31 / 95 & 51.55\% & \textbf{2.46} \\
\end{tabular}
\end{ruledtabular}
\label{tab:po_topology_rruff325}
\end{table*}

\subsection{Topological Evaluation of Symmetry Degradation}
\label{sec:topology}

Standard Top-1 accuracy treats every misclassification as an equal categorical failure, but crystallographically some errors are far more reasonable than others. We therefore mapped predictions onto the condensed directed acyclic graph (DAG) of maximal \textit{translationengleiche} ($t$-) subgroups---subgroups that preserve the translation lattice while removing point-group operations---linking the 99 extinction groups. One trigonal cycle is merged into a single node in this condensed graph (see Supplemental Material). In this graph, an edge represents the loss of a symmetry operation without changing the underlying translation lattice.

This analysis reveals a stark behavioral contrast. The older uncalibrated replay-era baseline (\textbf{Legacy}) has a mean topological error distance of 3.50 and is heavily biased toward ancestor hallucinations (158 higher-symmetry errors vs.\ 64 descendant errors), indicating reliance on chemical priors to guess high-symmetry structures while ignoring contradictory spectral evidence. By contrast, the calibrated non-PO Stage-2c model is much more local and conservative (Table~\ref{tab:po_topology_rruff325}): 38.4\% of its wrong Top-1 predictions lie within graph distance $\leq 2$ of the ground truth (mean distance 2.72), and its directed errors flow predominantly downward---191 descendant vs.\ 27 ancestor errors. The PO-informed ablations then reveal a more nuanced picture. Pure PO training sharply improves local accuracy (53.0\% of errors within $\leq 2$ hops) and reduces descendant errors to 157, but it also drives branch jumps upward and destabilizes split-head validity. The final large mixed model preserves most of that locality gain while recovering broad candidate coverage: 51.55\% of wrong predictions lie within graph distance $\leq 2$, the mean distance drops to 2.46, descendant errors fall to 165, and ancestor predictions rise modestly to 31 (Table~\ref{tab:po_topology_rruff325}).

This behavior is consistent with the physics of the problem. Higher-symmetry extinction groups mandate strict systematic absences (regions of zero intensity). Real-world background noise and impurity phases add intensity to reciprocal space, while preferred orientation can suppress whole reflection families. The non-PO model acts like a cautious crystallographer: observing intensity where an absence should be, it rejects the unsupported higher symmetry and conservatively defaults to a lower-symmetry descendant. Once explicit PO is introduced, the model learns that some missing reflections are not true absences but texture-suppressed observations. This weakens the descendant bias and tightens the local graph neighborhood of the remaining errors. At the same time, branch jumps remain elevated, which we interpret as a \emph{texture aliasing} effect: once preferred orientation erases a diagnostic family of reflections, the surviving 1D barcode can become locally ambiguous between nearby crystallographic cousins rather than merely lower-symmetry descendants.

This descendant-biased error pattern also mirrors human crystallographic workflow, in which practitioners often fall back to lower-symmetry extinction groups when the data do not unambiguously support a higher-symmetry assignment.

\begin{figure*}[t]
\centering
\includegraphics[width=0.90\textwidth]{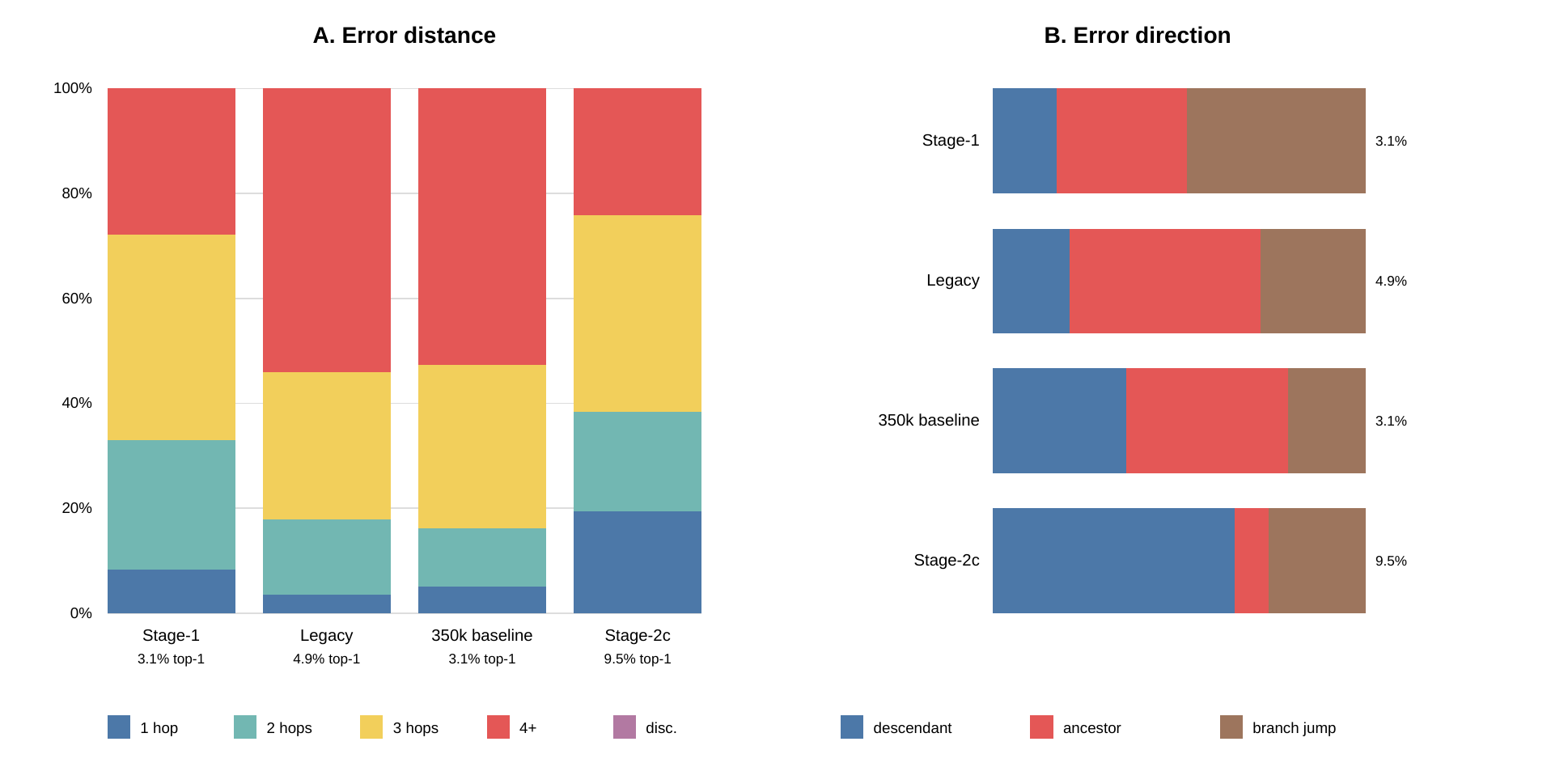}
\caption{Topological structure of Top-1 errors on the RRUFF-325 benchmark. Left: error-distance distribution on the condensed extinction-group subgroup graph. Right: directionality. Calibration and preferred-orientation-aware training shift errors toward shorter graph distances, reduce the strong descendant bias of the non-PO model, and reveal that many residual failures are local lateral hops rather than distant hallucinations.}
\label{fig:topological}
\end{figure*}

\subsection{The Catastrophic Paradox}

\begin{figure}[h]
\centering
\includegraphics[width=0.45\textwidth]{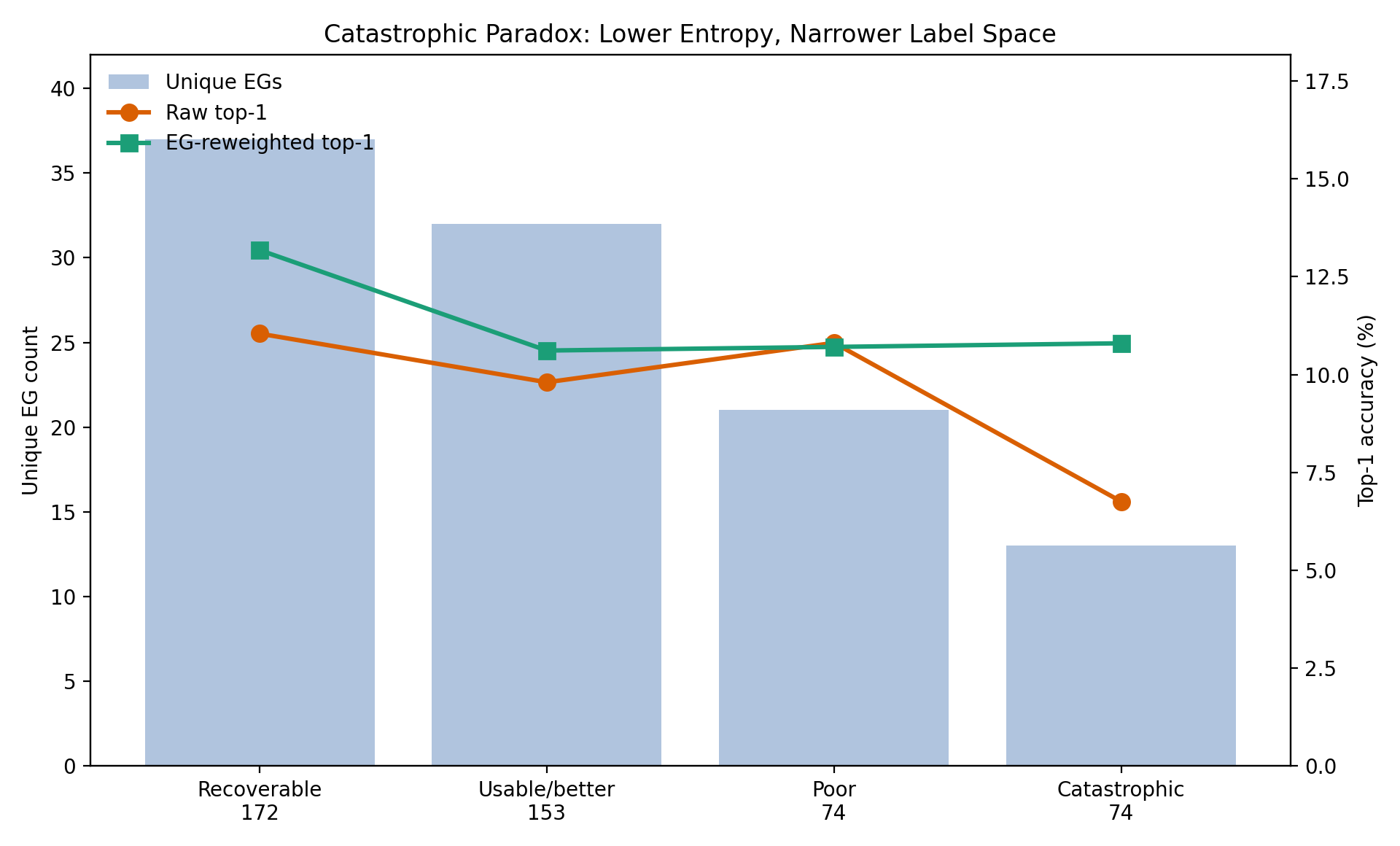}
\caption{The catastrophic paradox. Classical profile-fit quality and neural classification difficulty do not align monotonically; extinction-group reweighting removes about 45\% of the recoverable-versus-catastrophic gap.}
\label{fig:catastrophic}
\end{figure}

One of the most striking findings is that classical Rietveld fit quality does \textit{not} cleanly predict neural-network classification difficulty (Fig.~\ref{fig:catastrophic}). The 74 catastrophic-fit patterns span only 13 unique extinction groups and 38 minerals, whereas the 172 recoverable patterns span 37 extinction groups and 111 minerals. Importantly, the effect is not driven by a few dominant minerals: the most frequent mineral in the catastrophic stratum accounts for only 4 of 74 scans. On the final mixed checkpoint, raw Top-1 across the recoverable, usable-or-better, poor, and catastrophic strata is 11.05\%, 9.80\%, 10.81\%, and 6.76\%, respectively. Reweighting each stratum to a common extinction-group distribution changes these values to 13.18\%, 10.61\%, 10.71\%, and 10.80\%, showing that label-space concentration explains about 45\% of the recoverable-versus-catastrophic gap, but not all of it.

There is also a materials explanation. Many catastrophic patterns arise from strongly textured or highly cleavable minerals, where preferred orientation severely distorts relative intensities without removing the underlying Bragg-angle topology. While Rietveld-style nuisance fits collapse because the profile model expects powder-averaged intensities, our neural models appear to rely less on exact relative peak heights and can still recognize the invariant geometric barcode when the intensity envelope is badly distorted.

The same entropic logic governs the model's predictive ``sinks.'' The final RRUFF-325 comparison shows that the dominant measured sinks are EG~4 and EG~99 rather than a universal collapse into a single fallback group. Throughout, EG numbers refer to our internal extinction-group lookup-table indices, released with the code and Supplemental Material; they are identifiers rather than a separate international standard. When noise obscures fine peak-splitting, predictions concentrate into nearby low-constraint groups within the correct Bravais neighborhood---a structured, physically interpretable fallback.

\section{Comparison with Prior Work}

Having established the calibrated model's performance and failure modes on our curated benchmarks, we now situate these results relative to prior work.

The most direct comparison is with Schopmans \textit{et al.}~\cite{schopmans_neural_2023}, who trained ResNet models on large on-the-fly synthetic data (up to 261M diffractograms) for space-group classification, reporting approximately 25\% Top-1 on a hand-filtered set of 942 RRUFF patterns restricted to 145 space groups.

The two studies differ on nearly every axis, making direct numeric comparison misleading:
\begin{itemize}
\item \textit{Classification target}: they predict space groups (145 retained classes); we predict extinction groups (all 99 classes, no exclusions).
\item \textit{Evaluation set}: their 942-pattern set was hand-filtered; our RRUFF-473 benchmark is algorithmically curated and retains difficult patterns.
\item \textit{Data generation}: they use ICSD-informed synthesis with strong chemical priors; we use a physics-first curriculum beginning from uniform, chemically random data.
\item \textit{Training scale}: up to 261M diffractograms vs.\ our 5.14M.
\end{itemize}

On absolute Top-1, we do not yet match their reported $\sim$25\%. On ranking breadth, our best Top-5 of 52.22\% on RRUFF-473 is competitive given the data-scale difference. On our harshest RRUFF-325 benchmark, the final large mixed model reaches calibrated 10.46\% Top-1 and 43.69\% Top-5, reflecting the additional difficulty of strongly degraded mixtures without hand-filtering while still yielding broad candidate coverage.

A separate low-background 1222-pattern RRUFF holdout tells a complementary story. On this cleaner real-data set, the final large mixed model reached calibrated 10.07\% Top-1 and 45.34\% Top-5 in extinction-group space, while exact split-head validity rose to 65.30\%, compared with 47.38\% on the harsher RRUFF-325 benchmark. This indicates that the rule-based pathway is not simply broken on experimental data; rather, its strict symbolic decoding becomes substantially more usable as nuisance severity decreases, while the auxiliary head remains the more reliable path under stronger texture, overlap, and impurity corruption.

Cao \textit{et al.}~\cite{simxrd2025} recently introduced SimXRD-4M, a large-scale multiphysical simulator producing 4.07M diffractograms across 230 space groups, and reported strong transfer to experimental RRUFF data. Direct comparison is not straightforward for several reasons: the two studies use different classification targets (space groups vs.\ extinction groups), different RRUFF subsets (their $\sim$3000-pattern evaluation set vs.\ our algorithmically curated RRUFF-473/325), different intensity representations ($d$--$I$ vs.\ fixed $2\theta$ grid), and different normalization conventions. An exploratory cross-evaluation of our Stage-2c model on a locally reconstructed Cu-like RRUFF subset of 2738 patterns (see Supplemental Material) produced calibrated performance consistent with our main benchmarks (10.04\% Top-1, 45.73\% Top-5 in extinction-group space), but this is not a controlled head-to-head comparison. Because the two systems differ in benchmark definition, input representation ($d$--$I$ versus a fixed $2\theta$ grid), normalization conventions, and target taxonomy, a controlled comparison would require nontrivial harmonization and retraining that is beyond the scope of the present study.

Simple distribution-only baselines help contextualize these numbers. An ICSD-frequency prior---which never inspects a diffraction pattern---reaches about 9.7\% Top-1 but only 27.2\% Top-5 on RRUFF-like extinction labels. Our final large mixed model is only modestly above that prior on Top-1, but reaches 43.69\% Top-5 on RRUFF-325 and 50.74\% Top-5 on RRUFF-473, indicating that it is not merely reproducing a static class histogram: the model uses diffraction evidence to re-rank candidates on a per-pattern basis even when the strongest symmetry cues are partially destroyed by noise (see Supplemental Material for full analysis). Direct head-to-head comparison against recent larger-scale simulated-pretraining studies and multi-instrument experimental benchmarks remains future work, because the current study uses a different target taxonomy, harder algorithmically curated RRUFF subsets, and a single fixed Cu-K$\alpha$ laboratory geometry.

Recent generative approaches to PXRD analysis aim at the harder problem of full structure recovery. These methods are not directly comparable, but an extinction-group predictor of the type developed here could serve as an upstream constraint for such pipelines, reducing the combinatorial search space before structure generation.

\section{Discussion}

\subsection{Architecture Matters, But So Does Data}

The scaling study demonstrates that transformers substantially outperform CNNs on this task. Yet even the best transformer fails on real data without the right training curriculum. The RRUFF fine-tuning stage alone accounts for the majority of the real-data performance gain (Fig.~\ref{fig:curriculum}), underscoring that the synthetic-to-real domain gap is not merely a matter of model capacity.

\subsection{The Multi-Task Design Principle}

The dual-head architecture embodies a general principle for scientific machine learning: separate the interpretable physics-driven prediction from the robust statistical prediction, train them jointly so each regularizes the other, and fuse at inference. This is preferable to either a pure physics-rules approach (too brittle for noisy real data) or a pure black-box classifier (no interpretability, no protection against prior collapse).

\subsection{Curriculum as Domain Adaptation}

The training pathway can be understood as curriculum domain adaptation: uniform pretraining learns symmetry cues without geological frequency bias; RRUFF-conditioned fine-tuning adapts those representations to realistic nuisance structure; late preferred-orientation ablations show that subtractive noise must be represented explicitly; and calibrated Bayesian decoding restores empirical mineralogical prevalence at inference time without reintroducing uncontrolled overconfidence. The final large mixed run suggests that nuisance realism must itself be balanced: a small amount of exact PO sharpens ranking, but the best overall deployment model retains a majority of ordinary powder-like mixtures.

\subsection{Classical Baselines}

Alongside the neural-network work, we developed a classical statistical baseline using sparse Pawley fitting for extinction-group ranking. Even with physically motivated information criteria (AIC/BIC), the unconstrained full-sweep classical selector was unstable: lower-symmetry models overfit profile imperfections, and accidental chemical absences mimic symmetry absences. A topology-guided conditional benchmark---ranking only candidates within the neural model's bounded subgroup neighborhood---is both better posed and much faster in practice, converting a brittle multi-hour global sweep into bounded local verification that completes in tens of seconds to minutes on the supported branches.

In a broader supported non-monoclinic follow-up on 34 RRUFF-325 cases, all 34 bounded runs completed; 8/34 recovered the exact top-ranked space group, all in the rhombohedral branch. Primitive monoclinic remained the hardest regime, but after low-angle truncation, reflection clustering, and a nonnegative ridge/NNLS-style inner solve, a broader 61-case monoclinic follow-up completed 51 cases, showing that the bounded backend can be made practical even in the most difficult branch, albeit with local rather than yet exact rankings (see Supplemental Material).

\subsection{Limitations and Next Steps}

Our best balanced harsh-benchmark result is the final large mixed model at 10.46\% Top-1 and 43.69\% Top-5 on the real RRUFF-325 benchmark, while a pure-PO ablation reaches the sharper but narrower 13.54\% / 40.92\% operating point. This is still far from solved symmetry assignment, and three factors appear to bound the current system. First, there is substantial label ambiguity. Even perfectly simulated patterns from some extinction groups become nearly indistinguishable after peak overlap, preferred orientation, and impurity structure compress the available evidence; the prior-only baseline's 9.7\% Top-1 suggests that achievable Top-1 on this distribution may be substantially constrained even for a strong model. The model's main value on this benchmark is therefore not only exact Top-1, but also the large Top-5 gain over that frequency baseline.

Second, there is persistent simulator mismatch. Our synthetic generator now includes exact preferred orientation in the late ablation stage, and that intervention clearly changes the real-data error structure, but it still does not capture the full nuisance realism of field data, particularly specimen displacement, fluorescence, and related instrument-specific artifacts. The subgroup-DAG analysis shows that many of the remaining errors are local rather than random, suggesting that the model usually learns the correct crystallographic neighborhood but lacks sufficient nuisance fidelity to resolve it. In particular, the persistence of local branch jumps even after PO-aware training is consistent with a 1D ``texture aliasing'' limit, where suppression of whole reflection families makes nearby crystallographic cousins difficult to distinguish from a powder trace alone.

The empirical ceiling of $\sim$10\% Top-1 for extinction-group classification on degraded real mixtures has implications beyond the present task. Since extinction-group determination requires only the binary presence or absence of reflections, whereas full structure solution requires accurate continuous intensities, the difficulty we observe even at the binary level suggests that end-to-end structure recovery from 1D powder data under comparable nuisance conditions faces at least comparable difficulty. These observations motivate modular pipelines in which a calibrated symmetry predictor constrains the search space for downstream structure-solution methods, whether classical (Rietveld) or generative (diffusion models, GNNs), rather than attempting unconstrained end-to-end inversion.

The most immediate next steps are therefore bounded rather than purely scaling-focused: (i)~targeted improvements to nuisance realism rather than brute-force data scaling; (ii)~broader real-data validation of the matched SG$\rightarrow$EG control and related target-taxonomy comparisons; and (iii)~tighter benchmark-level analyses of how calibration and structured decoding fail under specific nuisance regimes. The main lesson of the stage-2 experiments is not simply ``more data helps'' but rather that calibration and nuisance realism determine whether the information already learned by the model becomes usable.

A late control supports this interpretation. Simple weight-space interpolation (WiSE-FT)~\cite{wortsman2022wiseft} between the broader Stage-2c checkpoint and the sharper PO-only checkpoint improved Top-1 relative to Stage-2c but did not recover the desired hybrid of PO-level Top-1 with Stage-2c-level Top-5. This suggests that the preferred-orientation effect is not merely a late linear interpolation in weight space, but instead depends on jointly learning texture and non-texture regimes during fine-tuning.

\section{Conclusions}

We have shown that reliable symmetry classification from powder diffraction requires three co-designed ingredients: the measurement-compatible target (extinction groups), a physics-informed architecture that separates crystallographic rule learning from statistical pattern matching, and a training curriculum that explicitly manages the synthetic-to-real domain gap.

Two results carry implications beyond this specific PXRD task. First, post-hoc calibration is not a cosmetic final step but a central part of the scientific inference pipeline: on degraded real data, temperature-scaled Bayesian decoding substantially improves deployment performance without retraining by decoupling learned structural evidence from the geological prior absorbed during fine-tuning. Second, the subgroup-DAG analysis shows that the calibrated model's residual errors are not random but physically structured. Preferred-orientation-aware training weakens the strong descendant bias of the non-PO model, while the final large mixed curriculum---combining standard and texture-augmented data---yields the tightest local topological error radius we observed. This graceful topological degradation emerges without explicit topological supervision, suggesting that calibrated inference can recover interpretable physical structure even from flexible neural classifiers.

These findings point toward a broader design pattern for scientific machine learning on spectroscopic and scattering data: encode the measurement physics in the architecture, let the training curriculum manage domain shift, and treat calibrated inference as a first-class design parameter rather than an afterthought. For powder diffraction specifically, the extinction-group framework and the RRUFF-473/RRUFF-325 benchmarks provide a foundation for community comparison, while the topology-guided classical verification pipeline shows how neural priors can convert an intractable global search into a bounded local problem. Attention is necessary but not sufficient: the physics must be in the architecture, the care must be in the data, and the final inference step must be calibrated to the deployment distribution.

\section*{Data availability}

Reproducibility materials for this study---including trained checkpoints, training and evaluation configurations, canonical evaluation wrappers, and the compact JSON artifacts underlying the benchmark, positional-ablation, calibration, and topological analyses reported here---are released through the public GitHub repository \url{https://github.com/scattering/paper-ai-diffraction} together with the Zenodo archival package \url{https://doi.org/10.5281/zenodo.19558452}. The release documents the required benchmark file names, expected local paths, and a reconstruction-oriented workflow for constructing the curated RRUFF-325 and RRUFF-473 benchmarks from the upstream RRUFF-derived sources.

\begin{acknowledgments}
The authors acknowledge the Texas Advanced Computing Center (TACC) at The University of Texas at Austin for resources made available to NIST under contract number 1333ND25PNB180410 that have contributed to the research results reported within this paper. URL: Texas Advanced Computing Center. Computational resources were also provided through ACCESS allocation PHY250007, ``Applications of AI to Diffraction.'' Support for Edward G. Friedman and Elizabeth Baggett was provided by the Center for High Resolution Neutron Scattering, a partnership between the National Institute of Standards and Technology and the National Science Foundation under Agreement No.~DMR-2010792. We thank Brian DeCost, Austin McDannald, Craig Brown, and Hui Wu for useful conversations. We thank the RRUFF project at the University of Arizona for making their mineral diffraction database publicly available.
\end{acknowledgments}

\clearpage
\onecolumngrid
\renewcommand{\thefigure}{S\arabic{figure}}
\renewcommand{\thetable}{S\arabic{table}}
\renewcommand{\thesection}{S\arabic{section}}
\renewcommand{\theequation}{S\arabic{equation}}
\setcounter{figure}{0}
\setcounter{table}{0}
\setcounter{section}{0}
\setcounter{equation}{0}
\section*{Supplemental Material}
\section{CNN Baseline Scaling}
\label{sec:cnn_scaling}

\begin{figure}[H]
\centering
\includegraphics[width=0.7\textwidth]{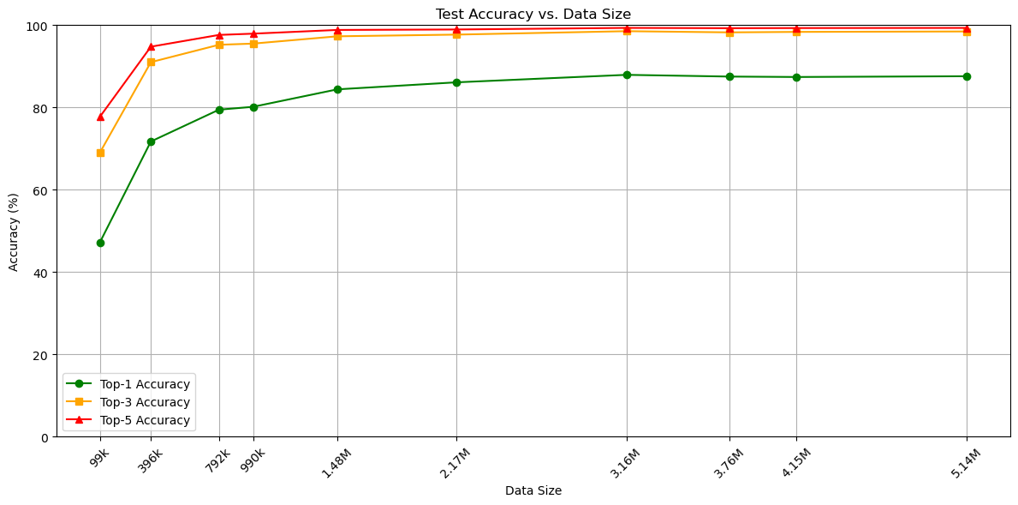}
\caption{Scaling ResNet-18 with synthetic reflection data. Performance plateaus beyond approximately 1M samples, in contrast to the continued improvement observed for transformers at the same data scales (see main-text Table~III).}
\label{fig:resnet_scaling}
\end{figure}

\section{Bias and Cross-Distribution Studies}
\label{sec:bias}

\begin{table}[H]
\caption{Cross-evaluation of biased (ICSD-distributed) and balanced ResNet-18 models. The balanced model generalizes substantially better to the opposite distribution.}
\begin{ruledtabular}
\begin{tabular}{lccc}
Evaluation & Top-1 & Top-3 & Top-5 \\
\colrule
Biased model on biased test     & 73.45\% & 92.16\% & 95.86\% \\
Balanced model on balanced test & 57.09\% & 81.31\% & 89.30\% \\
\colrule
Biased model on balanced test   & 38.68\% & 62.85\% & 76.19\% \\
Balanced model on biased test   & 54.00\% & 81.26\% & 89.75\% \\
\end{tabular}
\end{ruledtabular}
\label{tab:cross_eval}
\end{table}

\section{Extinction-Group Mapping Table}
\label{sec:eg_table}

For reproducibility, Table~\ref{tab:full_eg_map} lists the full 99-extinction-group mapping used throughout this work. Each extinction group is identified by the same integer label used in training and evaluation, together with a canonical Hermann--Mauguin-style representative and the associated crystallographic space-group numbers.

\small
\newcommand{\egtrow}[3]{#1 & \parbox[t]{0.62\textwidth}{\raggedright #2} & \parbox[t]{0.22\textwidth}{\raggedright #3} \\}
\begin{longtable}{lll}
\caption{Full extinction-group lookup used in the 99-class formulation. Canonical labels are taken from the code-level lookup table, and the associated space-group numbers are the complete crystallographically indistinguishable sets under the powder-diffraction extinction conditions used here.}
\label{tab:full_eg_map}\\
\toprule
\parbox[t]{0.06\textwidth}{EG} & \parbox[t]{0.62\textwidth}{\raggedright Canonical extinction group} & \parbox[t]{0.22\textwidth}{\raggedright Space-group numbers} \\
\midrule
\endfirsthead
\toprule
\parbox[t]{0.06\textwidth}{EG} & \parbox[t]{0.62\textwidth}{\raggedright Canonical extinction group} & \parbox[t]{0.22\textwidth}{\raggedright Space-group numbers} \\
\midrule
\endhead
\midrule
\multicolumn{3}{r}{Continued on next page}\\
\endfoot
\bottomrule
\endlastfoot
\egtrow{1}{P - 1 1 (equiv: P 1 - 1, P 1 1 -)}{3, 6, 10}
\egtrow{2}{P 21 1 1 (equiv: P 1 21 1, P 1 1 21)}{4, 11}
\egtrow{3}{P b 1 1 (equiv: P c 1 1, P n 1 1, P 1 a 1, P 1 c 1, P 1 n 1, P 1 1 a, P 1 1 b, P 1 1 n)}{7, 13}
\egtrow{4}{P 21/b 1 1 (equiv: P 21/c 1 1, P 21/n 1 1, P 1 21/a 1, P 1 21/c 1, P 1 21/n 1, P 1 1 21/a, P 1 1 21/b, P 1 1 21/n)}{14}
\egtrow{5}{C - 1 1 (equiv: B - 1 1, I - 1 1, C 1 - 1, A 1 - 1, I 1 - 1, B 1 1 -, A 1 1 -, I 1 1 -)}{5, 8, 12}
\egtrow{6}{C n 1 1 (equiv: B b 1 1, I c 1 1, C 1 c 1, A 1 n 1, I 1 a 1, B 1 1 n, A 1 1 a, I 1 1 b)}{9, 15}
\egtrow{7}{P - - -}{16, 25, 47}
\egtrow{8}{P - - 21 (equiv: P - 21 -, P 21 - -)}{17}
\egtrow{9}{P - 21 21 (equiv: P 21 - 21, P 21 21 -)}{18}
\egtrow{10}{P 21 21 21}{19}
\egtrow{11}{P - - a (equiv: P - - b, P - a -, P - c -, P b - -, P c - -)}{26, 28, 51}
\egtrow{12}{P - - n (equiv: P - n -, P n - -)}{31, 59}
\egtrow{13}{P - a a (equiv: P b - b, P c c -)}{27, 49}
\egtrow{14}{P - a b (equiv: P - c a, P b - a, P b c -, P c - b, P c a -)}{29, 57}
\egtrow{15}{P - a n (equiv: P - n a, P b - n, P c n -, P n - b, P n c -)}{30, 53}
\egtrow{16}{P - c b (equiv: P b a -, P c - a)}{32, 55}
\egtrow{17}{P - c n (equiv: P - n b, P b n -, P c - n, P n - a, P n a -)}{33, 62}
\egtrow{18}{P - n n (equiv: P n - n, P n n -)}{34, 58}
\egtrow{19}{P b a a (equiv: P b a b, P b c b, P c a a, P c c a, P c c b)}{54}
\egtrow{20}{P b a n (equiv: P c n a, P n c b)}{50}
\egtrow{21}{P b c a (equiv: P c a b)}{61}
\egtrow{22}{P b c n (equiv: P b n a, P c a n, P c n b, P n a b, P n c a)}{60}
\egtrow{23}{P b n b (equiv: P c c n, P n a a)}{56}
\egtrow{24}{P b n n (equiv: P c n n, P n a n, P n c n, P n n a, P n n b)}{52}
\egtrow{25}{P n n n}{48}
\egtrow{26}{C - - - (equiv: B - - -, A - - -)}{21, 35, 38, 65}
\egtrow{27}{C - - 21 (equiv: B - 21 -, A 21 - -)}{20}
\egtrow{28}{C - - (ab) (equiv: B - (ac)-, A(bc)- -)}{39, 67}
\egtrow{29}{C - c - (equiv: C c - -, B - - b, B b - -, A - - a, A - a -)}{36, 40, 63}
\egtrow{30}{C - c (ab) (equiv: C c - (ab), B - (ac)b, B b (ac)-, A(bc)- a, A(bc)a -)}{41, 64}
\egtrow{31}{C c c - (equiv: B b - b, A - a a)}{37, 66}
\egtrow{32}{C c c (ab) (equiv: B b (ac)b, A(bc)a a)}{68}
\egtrow{33}{I - - -}{23, 24, 44, 71}
\egtrow{34}{I - - (ab) (equiv: I - (ac)-, I(bc)- -)}{46, 74}
\egtrow{35}{I - c b (equiv: I c - a, I b a -)}{45, 72}
\egtrow{36}{I b c a}{73}
\egtrow{37}{F - - -}{22, 42, 69}
\egtrow{38}{F - d d (equiv: F d - d, F d d -)}{43}
\egtrow{39}{F d d d}{70}
\egtrow{40}{P - - -}{75, 81, 83, 89, 99, 111, 115, 123}
\egtrow{41}{P - 21 -}{90, 113}
\egtrow{42}{P 42 - -}{77, 84, 93}
\egtrow{43}{P 42 21 -}{94}
\egtrow{44}{P 41 - -}{76, 78, 91, 95}
\egtrow{45}{P 41 21 -}{92, 96}
\egtrow{46}{P - - c}{105, 112, 131}
\egtrow{47}{P - 21 c}{114}
\egtrow{48}{P - b -}{100, 117, 127}
\egtrow{49}{P - b c}{106, 135}
\egtrow{50}{P - c -}{101, 116, 132}
\egtrow{51}{P - c c}{103, 124}
\egtrow{52}{P - n -}{102, 118, 136}
\egtrow{53}{P - n c}{104, 128}
\egtrow{54}{P n - -}{85, 129}
\egtrow{55}{P 42/n - -}{86}
\egtrow{56}{P n - c}{137}
\egtrow{57}{P n b -}{125}
\egtrow{58}{P n b c}{133}
\egtrow{59}{P n c -}{138}
\egtrow{60}{P n c c}{130}
\egtrow{61}{P n n -}{134}
\egtrow{62}{P n n c}{126}
\egtrow{63}{I - - -}{79, 82, 87, 97, 107, 119, 121, 139}
\egtrow{64}{I 41 - -}{80, 98}
\egtrow{65}{I - - d}{109, 122}
\egtrow{66}{I - c -}{108, 120, 140}
\egtrow{67}{I - c d}{110}
\egtrow{68}{I 41/a - -}{88}
\egtrow{69}{I a - d}{141}
\egtrow{70}{I a c d}{142}
\egtrow{71}{P - - -}{143, 147, 149, 150, 156, 157, 162, 164, 168, 174, 175, 177, 183, 187, 189, 191}
\egtrow{72}{P 31 - -}{144, 145, 151, 152, 153, 154}
\egtrow{73}{P - - c}{159, 163, 186, 190, 194}
\egtrow{74}{P - c -}{158, 165, 185, 188, 193}
\egtrow{75}{R (obv) - - (equiv: R (rev) - -, R - - -)}{146, 148, 155, 160, 166}
\egtrow{76}{R (obv)- - c (equiv: R (rev)- - c, R - - c)}{161, 167}
\egtrow{77}{P 63 - -}{173, 176, 182}
\egtrow{78}{P 62 - -}{171, 172, 180, 181}
\egtrow{79}{P 61 - -}{169, 170, 178, 179}
\egtrow{80}{P - c c}{184, 192}
\egtrow{81}{P - - -}{195, 200, 207, 215, 221}
\egtrow{82}{P 21 - -}{198}
\egtrow{83}{P 42 - -}{208}
\egtrow{84}{P 41 - -}{212, 213}
\egtrow{85}{P - - n}{218, 223}
\egtrow{86}{P a - -}{205}
\egtrow{87}{P n - -}{201, 224}
\egtrow{88}{P n - n}{222}
\egtrow{89}{I - - -}{197, 199, 204, 211, 217, 229}
\egtrow{90}{I 41 - -}{214}
\egtrow{91}{I - - d}{220}
\egtrow{92}{I a - -}{206}
\egtrow{93}{I a - d}{230}
\egtrow{94}{F - - -}{196, 202, 209, 216, 225}
\egtrow{95}{F 41 - -}{210}
\egtrow{96}{F - - c}{219, 226}
\egtrow{97}{F d - -}{203, 227}
\egtrow{98}{F d - c}{228}
\egtrow{99}{P -}{1, 2}
\end{longtable}
\let\egtrow\relax
\normalsize

\begin{table}[H]
\caption{Bias cross-evaluation for the vision transformer. The pattern is consistent: balanced training generalizes better to biased evaluation than vice versa.}
\begin{ruledtabular}
\begin{tabular}{lccc}
Model & Top-1 & Top-3 & Top-5 \\
\colrule
VT -- Biased Reflection     & 78.73\% & 92.79\% & 96.44\% \\
VT -- Biased PyXtal          & 53.78\% & 76.66\% & 84.45\% \\
\colrule
VT -- Balanced Refl.\ on Biased Test & 87.06\% & 97.93\% & 99.25\% \\
VT -- Biased Refl.\ on Balanced Test & 44.35\% & 68.18\% & 78.35\% \\
VT -- Balanced PyXtal on Biased Test     & 50.49\% & 75.44\% & 84.51\% \\
VT -- Biased PyXtal on Balanced Test     & 22.01\% & 37.68\% & 47.41\% \\
\end{tabular}
\end{ruledtabular}
\label{tab:vt_cross_eval}
\end{table}

\section{Ablation: Data Generation Method and $2\theta$ Range}
\label{sec:ablation}

\begin{table}[H]
\caption{Vision transformer ablation on generation method and $2\theta$ range (synthetic test sets). Reflection-based data outperforms PyXtal-based data by a large margin.}
\begin{ruledtabular}
\begin{tabular}{lccc}
Model & Top-1 & Top-3 & Top-5 \\
\colrule
VT -- Reduced Reflection (10--80\textdegree) & 88.52\% & 98.53\% & 99.44\% \\
VT -- Reduced PyXtal (10--80\textdegree)     & 51.46\% & 75.01\% & 83.87\% \\
\end{tabular}
\end{ruledtabular}
\label{tab:ablation_generation}
\end{table}

\begin{table}[H]
\caption{Performance on real RRUFF data for different generation methods and $2\theta$ ranges. Reduced range is 10--80\textdegree; full range is 10--110\textdegree. The regular transformer benefits slightly from the wider range on reflection data; the vision transformer is less sensitive to range changes.}
\begin{ruledtabular}
\begin{tabular}{lccc}
Test Configuration & Top-1 & Top-3 & Top-5 \\
\colrule
RT -- Reduced Refl.\ on Reduced RRUFF & 8.94\% & 19.53\% & 28.62\% \\
RT -- Full Refl.\ on Full RRUFF       & 9.98\% & 21.29\% & 25.69\% \\
RT -- Reduced PyXtal on Reduced RRUFF     & 4.15\% & 10.45\% & 18.50\% \\
RT -- Full PyXtal on Full RRUFF           & 1.93\% & 7.69\%  & 11.13\% \\
\colrule
VT -- Reduced Refl.\ on Reduced RRUFF & 7.65\% & 14.13\% & 20.21\% \\
VT -- Full Refl.\ on Full RRUFF       & 7.41\% & 14.53\% & 19.93\% \\
VT -- Reduced PyXtal on Reduced RRUFF     & 7.98\% & 15.13\% & 19.50\% \\
VT -- Full PyXtal on Full RRUFF           & 9.02\% & 14.92\% & 23.26\% \\
\end{tabular}
\end{ruledtabular}
\label{tab:real_ablation}
\end{table}

\section{Training-Distribution Matrices (10M PyXtal Dataset)}
\label{sec:distribution_matrices}

These tables report Top-1 accuracy when models trained on one label distribution are evaluated on test sets drawn from a different distribution.

\begin{table}[H]
\caption{Training-distribution effect on regular transformer performance (10M PyXtal dataset, Top-1). Models trained on a specific distribution perform best on that same distribution but generalize poorly---especially to augmented (RRUFF-distribution) test data.}
\begin{ruledtabular}
\begin{tabular}{lcccc}
RT Train $\downarrow$ / Test $\rightarrow$ & Balanced & ICSD & RRUFF & Augmented \\
\colrule
Balanced  & 45.25\% & 41.04\% & 43.97\% & 0.86\% \\
ICSD      & 21.29\% & 61.29\% & 57.83\% & 9.93\% \\
RRUFF     & 19.88\% & 53.31\% & 60.08\% & 7.88\% \\
Augmented & 9.11\%  & 31.79\% & 36.88\% & 36.73\% \\
\end{tabular}
\end{ruledtabular}
\label{tab:rt_distribution}
\end{table}

\begin{table}[H]
\caption{Training-distribution effect on vision transformer performance (10M PyXtal dataset, Top-1). Same pattern as the regular transformer.}
\begin{ruledtabular}
\begin{tabular}{lcccc}
VT Train $\downarrow$ / Test $\rightarrow$ & Balanced & ICSD & RRUFF & Augmented \\
\colrule
Balanced  & 60.75\% & 59.66\% & 62.64\% & 3.86\% \\
ICSD      & 42.12\% & 74.34\% & 72.48\% & 1.84\% \\
RRUFF     & 38.43\% & 69.10\% & 76.25\% & 1.99\% \\
Augmented & 1.67\%  & 12.99\% & 10.83\% & 34.84\% \\
\end{tabular}
\end{ruledtabular}
\label{tab:vt_distribution}
\end{table}

\begin{table}[H]
\caption{Training-distribution effect on CNN performance (10M PyXtal dataset, Top-1). Training Data on RRUFF Distribution not tested on CNN}
\begin{ruledtabular}
\begin{tabular}{lcccc}
CNN Train $\downarrow$ / Test $\rightarrow$ & Balanced & ICSD & RRUFF & Augmented \\
\colrule
Balanced  & 56.1\% & 51.2\% & --- & 0.7\% \\
ICSD      & 38.7\% & 73.5\% & --- & 3.2\% \\
Augmented & 1.2\%  & 2.1\%  & --- & 34.6\% \\
\end{tabular}
\end{ruledtabular}
\label{tab:cnn_distribution}
\end{table}

\section{Effect of ICSD Duplicates}
\label{sec:duplicates}

\begin{table}[H]
\caption{Effect of removing duplicate entries from the ICSD on ResNet-18 classification performance. The drop is not solely attributable to reduced dataset size (see Fig.~\ref{fig:resnet_scaling}), suggesting that duplicates provide a form of implicit data augmentation.}
\begin{ruledtabular}
\begin{tabular}{lccc}
Dataset & Top-1 & Top-3 & Top-5 \\
\colrule
ICSD with duplicates (180k)    & 66\% & 80\% & 84\% \\
ICSD without duplicates (120k) & 55\% & 71\% & 78\% \\
\end{tabular}
\end{ruledtabular}
\label{tab:duplicates}
\end{table}

\section{Balanced Model Performance on Top ICSD Extinction Groups}
\label{sec:top10}

\begin{figure}[H]
\centering
\includegraphics[width=0.7\textwidth]{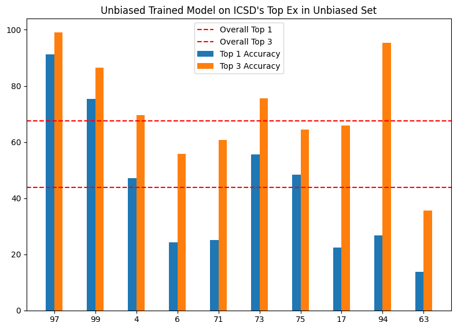}
\caption{Performance of the balanced ResNet-18 model on the ten most common extinction groups in the ICSD. For approximately half of these classes, the model's accuracy is markedly lower than its overall average, reflecting the difficulty of distinguishing high-population extinction groups that share similar structural motifs.}
\label{fig:top10_performance}
\end{figure}

\section{Confusion Matrix}
\label{sec:confusion}

\begin{figure}[H]
\centering
\includegraphics[width=0.8\textwidth]{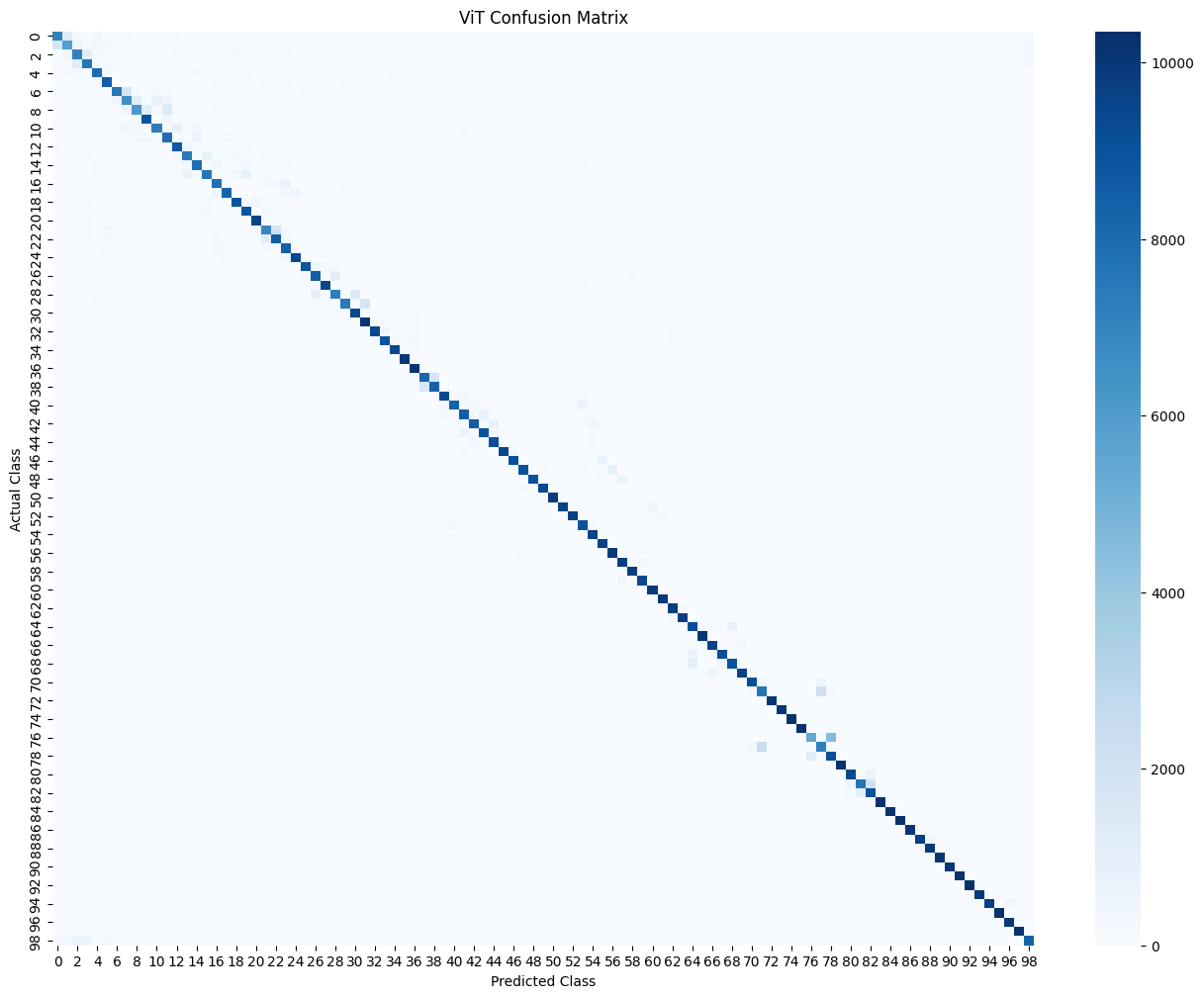}
\caption{Confusion matrix for the vision transformer on the synthetic balanced test set. Off-diagonal concentrations correspond to extinction groups that are crystallographic near-neighbors.}
\label{fig:confusion_vit}
\end{figure}

\section{Attention Map Visualizations}
\label{sec:attention_maps}

Attention-map visualizations are useful as qualitative diagnostics but are not direct proofs of mechanism~\cite{wiegreffe2019attentionexplanation}. In our architecture comparisons, both transformer variants distributed attention across minor peaks as well as dominant peaks, whereas CNN saliency concentrated more strongly on the tallest reflections. This behavior is qualitatively consistent with the idea that extinction-group discrimination depends on weak systematic-absence cues that CNNs may underweight~\cite{D3DD00198A}. Because these visualizations come from the earlier architecture study rather than the final calibrated stage-2 checkpoint, we treat them as suggestive only.

\begin{figure}[H]
\centering
\includegraphics[width=0.7\textwidth]{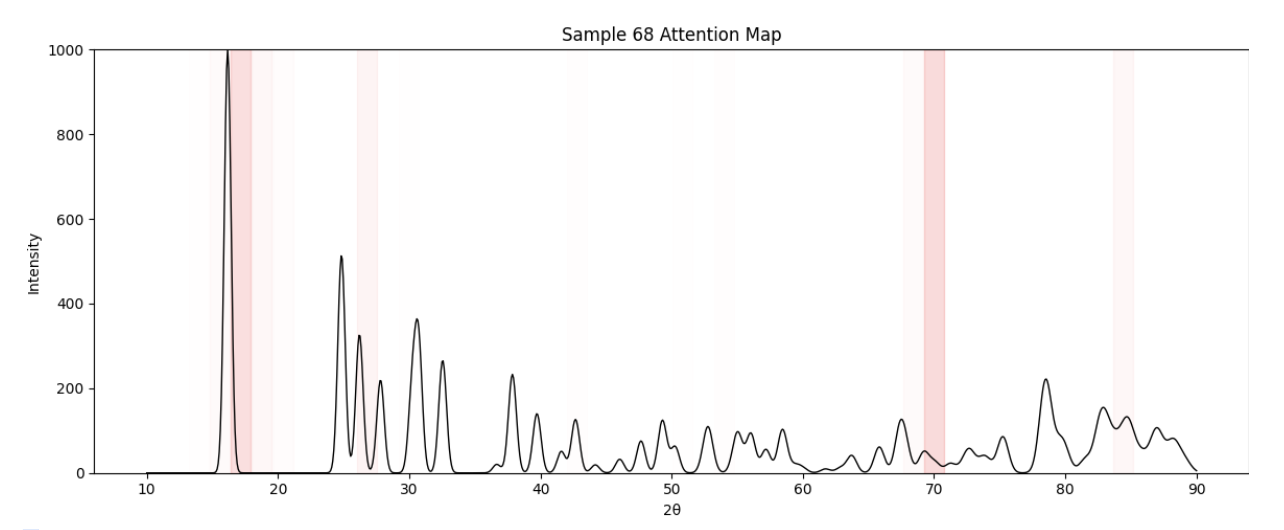}
\caption{Regular transformer attention map. Attention concentrates on both primary and minor peaks, consistent with the model learning systematic-absence rules.}
\label{fig:rt_attention}
\end{figure}

\begin{figure}[H]
\centering
\includegraphics[width=0.7\textwidth]{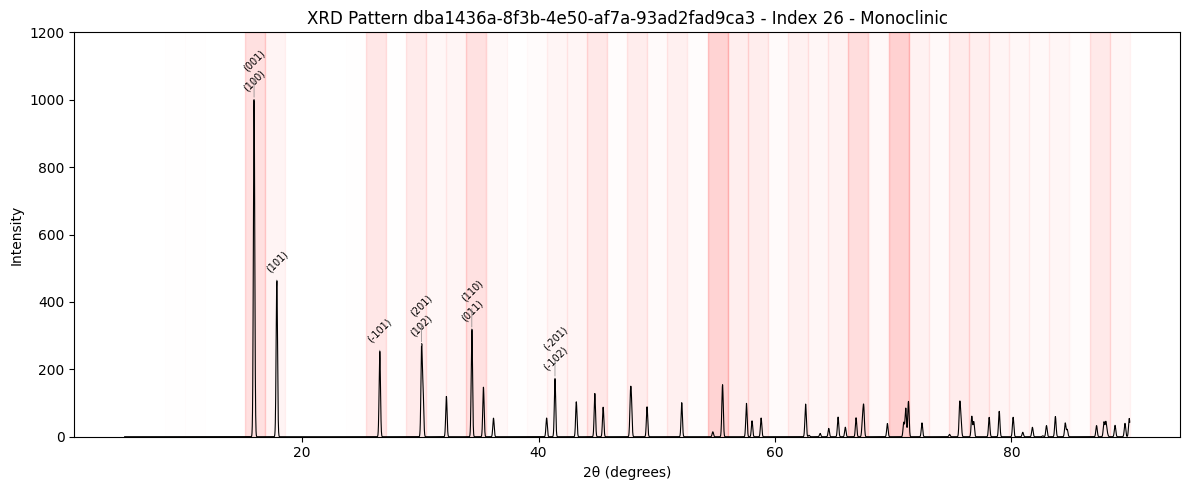}
\caption{Vision transformer attention map. Strong attention in the low-angle region (hkl 001/100/101) and primary peaks, with diffuse attention at high angles.}
\label{fig:vt_attention}
\end{figure}

\begin{figure}[H]
\centering
\includegraphics[width=0.7\textwidth]{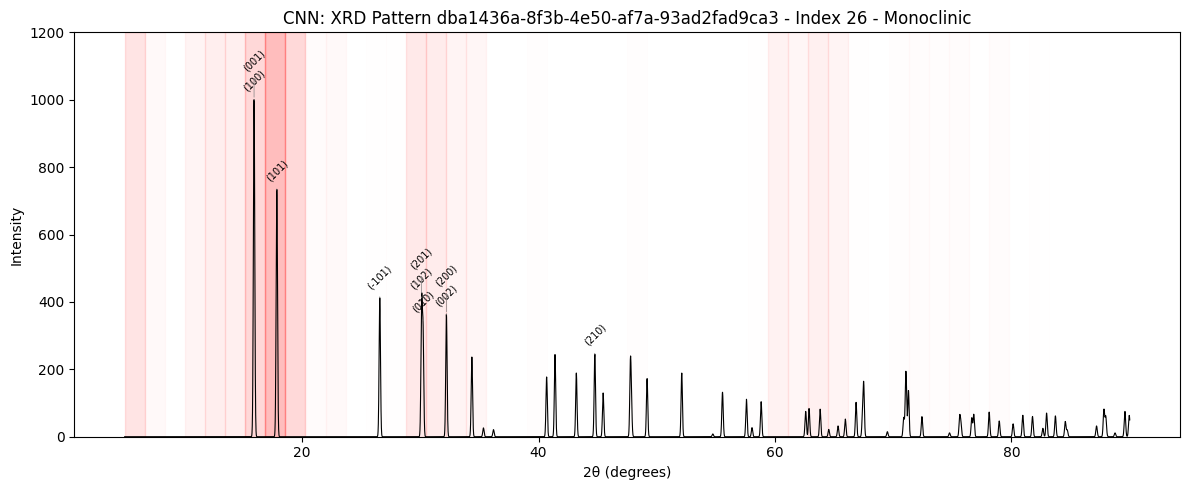}
\caption{CNN saliency map (Grad-CAM). The CNN attends primarily to the tallest peaks, with less sensitivity to the minor peaks that encode symmetry information.}
\label{fig:cnn_saliency}
\end{figure}

We also generated a small set of overlays from the final calibrated Stage-2c checkpoint on six RRUFF-325 cases: three correct predictions and three descendant-error cases. The updated overlays are more selective than the earlier architecture-study examples and include HKL annotations restricted to high-attention windows. However, both visual inspection and a lightweight quantitative analysis indicate that these maps remain better as diagnostics than as mechanistic evidence. Attention is still broadly distributed outside narrow observed-peak windows, and the contrast between correct and descendant-error cases is modest rather than decisive.

\begin{figure}[H]
\centering
\includegraphics[width=\textwidth]{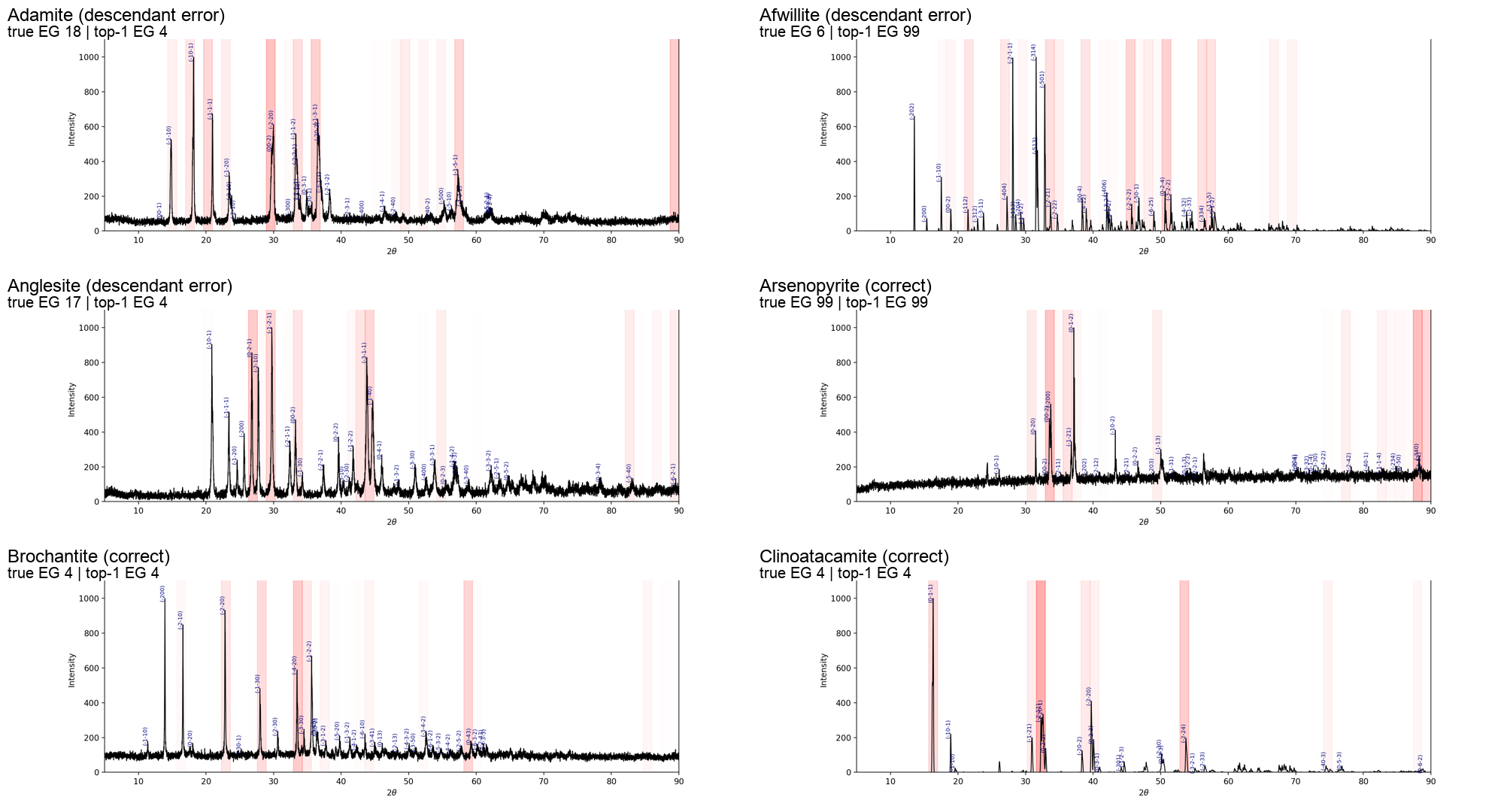}
\caption{Six calibrated Stage-2c attention overlays on RRUFF-325: three correct cases (top) and three descendant-error cases (bottom). Labels are restricted to well-separated peaks within high-attention windows. A small quantitative follow-up on this panel shows that descendant-error cases place somewhat more attention mass in inter-peak / absence-like windows than correct cases, but the effect is still modest. We therefore treat these maps as supplementary interpretability diagnostics rather than as direct proof of systematic-absence reasoning.}
\label{fig:stage2c_attention_panel}
\end{figure}

\section{Benchmark-Level Failure Analysis and Topological Error Structure}
\label{sec:failure_analysis}

The RRUFF-473 benchmark exposes failure modes that are largely invisible in aggregate Top-$k$ metrics. In the recoverable stratum, misclassifications are not dominated by single-step near-neighbor confusions alone; instead, predictions collapse disproportionately into a small set of sink extinction groups. In our earlier internal analyses EG~98 looked like a plausible universal attractor, but the final calibrated RRUFF-325 comparison shows that the dominant sinks of the current champion are actually EG~4 and EG~99. The correct conclusion is therefore not that one fixed low-assumption extinction group explains all failures, but that the calibrated model's errors remain structured and concentrated rather than diffuse across the full label space.

We analyzed these failures on a condensed extinction-group subgroup graph induced from maximal \textit{translationengleiche} subgroup relations between space groups. This graph has 98 nodes after condensing the single trigonal cycle between the two swapped $P3c1/P31c$-type extinction classes into one node. On RRUFF-325, the calibrated Stage-2c checkpoint places 38.4\% of its wrong Top-1 predictions within graph distance $\leq 2$ of the ground truth, compared with only 17.8\% for the older Legacy auxiliary baseline (Table~\ref{tab:topological_summary}). More importantly, the calibrated model's directed errors are strongly asymmetric: 191 wrong predictions move downward to descendant (lower-symmetry) nodes, versus only 27 upward ancestor hallucinations.

We therefore interpret the calibrated model's residual errors as \emph{graceful topological degradation}: when noise destroys the evidence for a specific screw or glide, the model tends to fall to nearby lower-symmetry alternatives rather than jumping arbitrarily across the crystallographic hierarchy. This is consistent with the physics of the problem: higher-symmetry extinction groups mandate strict systematic absences (regions of zero intensity), and real-world background, preferred orientation, and impurity phases add intensity that physically obscures these zero-intensity cues.

\begin{table}[H]
\caption{Topological error summary on the harsh RRUFF-325 benchmark. The calibrated champion exhibits shorter mean error distance, more errors within 2 hops, and strongly asymmetric descendant-biased directionality.}
\begin{ruledtabular}
\begin{tabular}{lcccc}
Model & Top-1 & Mean dist. & $\leq$2 hops & Desc./Anc. \\
\colrule
Stage-1 split & 3.08\% & 2.93 & 33.0\% & 54/110 \\
Legacy aux ($T\!=\!1$) & 4.92\% & 3.50 & 17.8\% & 64/158 \\
350k baseline aux ($T\!=\!1$) & 3.08\% & 3.47 & 16.2\% & 113/137 \\
Stage-2c aux ($T\!=\!5$) & 9.54\% & 2.72 & 38.4\% & 191/27 \\
\end{tabular}
\end{ruledtabular}
\label{tab:topological_summary}
\end{table}

The ``catastrophic paradox'' is also visible in the label statistics. The 74 catastrophic scans span only 13 extinction groups and 38 minerals, whereas the 172 recoverable scans span 37 extinction groups and 111 minerals. Thus, even though catastrophic patterns are worse from the standpoint of nuisance fitting, they present a narrower classification problem in extinction-group space. We therefore treat nuisance-fit quality and label-space entropy as distinct axes of evaluation rather than a single notion of data quality.

\begin{table}[H]
\caption{Controlled view of the catastrophic paradox on RRUFF-473 for the final mixed checkpoint. Reweighting by the extinction-group histogram removes about 45\% of the recoverable-versus-catastrophic gap, showing that much of the effect comes from label-space concentration.}
\begin{ruledtabular}
\begin{tabular}{lccccc}
Severity stratum & $n$ & Raw Top-1 (\%) & EG-reweighted Top-1 (\%) & Unique EGs & Unique minerals \\
\colrule
Catastrophic & 74 & 6.76 & 10.80 & 13 & 38 \\
Poor & 74 & 10.81 & 10.71 & 21 & 57 \\
Recoverable & 172 & 11.05 & 13.18 & 37 & 111 \\
Usable or better & 153 & 9.80 & 10.61 & 32 & 90 \\
\end{tabular}
\end{ruledtabular}
\label{tab:catastrophic_control}
\end{table}

This control sharpens the interpretation. Lower label-space entropy explains much of the effect: once the strata are reweighted to a common extinction-group composition, about 45\% of the recoverable-versus-catastrophic gap disappears. The remaining difference is therefore consistent with genuine profile-fit mismatch, not only with label-distribution bias.

\section{Split-Head Component Accuracy on Real Benchmarks}
\label{sec:split_components}

The 37-bit split-head target is a deterministic crystallographic encoding, not a latent representation interpreted only after training. Seven bits encode crystal system, five bits encode lattice centering, and the remaining 25 bits encode screw/glide operators in the first, second, and third symbol positions of the canonical extinction-group notation. Each extinction group is first assigned a canonical Hermann--Mauguin-style representative, and that symbol is then parsed deterministically into a fixed binary target vector. The first token specifies lattice type (e.g.\ $P$, $I$, $F$, $C$, $R$), while subsequent tokens specify the presence of position-specific operators such as $2_1$ screws and $a,b,c,d,n,ab$ glides. The split head therefore predicts an explicitly defined rule vector rather than an emergent internal code.

To separate coarse geometric localization from brittle exact extinction-group decoding, we evaluated the split head directly on the real RRUFF-473 and RRUFF-325 benchmarks. Table~\ref{tab:split_components_real} reports crystal-system accuracy, lattice-centering accuracy, operator-bit accuracy, and exact operator-match accuracy for the main curriculum checkpoints.

\begin{table}[H]
\caption{Direct split-head component accuracy (\%) on the real RRUFF-473 and RRUFF-325 benchmarks. Stage-2 fine-tuning improves operator-bit fidelity but degrades lattice-centering accuracy relative to the Stage-1 checkpoint, while exact operator-rule recovery remains very brittle on measured data.}
\begin{ruledtabular}
\begin{tabular}{llcccc}
Benchmark & Checkpoint & Crystal system & Lattice & Operator bit & Operator exact match \\
\colrule
RRUFF-325 & Stage-1 & 32.00 & 65.54 & 73.98 & 0.00 \\
RRUFF-325 & Stage-2c & 32.00 & 38.46 & 82.50 & 0.62 \\
RRUFF-325 & Stage-2c-aux025 & 32.92 & 39.69 & 82.14 & 1.23 \\
\colrule
RRUFF-473 & Stage-1 & 34.46 & 63.64 & 74.60 & 0.00 \\
RRUFF-473 & Stage-2c & 35.94 & 40.38 & 82.17 & 0.63 \\
RRUFF-473 & Stage-2c-aux025 & 36.15 & 41.01 & 81.78 & 1.06 \\
\end{tabular}
\end{ruledtabular}
\label{tab:split_components_real}
\end{table}

These component metrics sharpen the decoder interpretation in the main text. First, exact symbolic rule recovery is extremely brittle on measured data: even after Stage-2 fine-tuning, exact operator-match accuracy remains near $1\%$. Second, Stage-2 fine-tuning does not materially improve crystal-system accuracy, but it does improve operator-bit accuracy by roughly eight percentage points. Third, that gain comes with a large drop in lattice-centering accuracy relative to the Stage-1 checkpoint. In other words, the structured path learns more local extinction logic, but this does not translate into robust exact symbolic decoding on noisy real mixtures. This is consistent with the main result that the calibrated auxiliary path, rather than the direct split path, remains the strongest deployment operating point on RRUFF-325.

\begin{table}[H]
\caption{Exact split-head validity on the real benchmarks for the final Stage-2c checkpoint. Here we evaluate the raw symbolic bit pattern itself: whether the argmax crystal-system class, argmax lattice class, and thresholded operator bits correspond to exactly one extinction-group template, to multiple templates, or to none at all.}
\begin{ruledtabular}
\begin{tabular}{lcccc}
Benchmark & Valid single-EG & Ambiguous multi-EG & Invalid zero-match & Correct within all samples \\
\colrule
RRUFF-325 & 6.15\% & 4.62\% & 89.23\% & 0.92\% \\
RRUFF-473 & 6.98\% & 4.23\% & 88.79\% & 0.85\% \\
\end{tabular}
\end{ruledtabular}
\label{tab:split_validity_real}
\end{table}

This analysis answers the decoding-consistency question directly. On real mixtures, the symbolic split head rarely emits a bit pattern that maps cleanly onto a unique extinction-group template, and most thresholded bit patterns are physically invalid under the full template bank. This is precisely why the split path is useful in our final system mainly as a structural regularizer and fusion component rather than as a standalone deployment decoder.

\section{Bayesian Prior Provenance}
\label{sec:prior_provenance}

The geological prior used for Bayesian auxiliary-head decoding is an Empirical Bayes prior estimated from the training corpus only. No labels from the real evaluation benchmark are used when constructing this prior. The goal is to restore the strongly non-uniform mineralogical prevalence of Earth materials after symmetry-balanced pretraining, not to tune the model to the benchmark itself.

The post-hoc temperature parameter is a separate matter. The auxiliary-head temperature sweep reported in the main text and in Sec.~\ref{sec:temperature} was performed directly on the harsh RRUFF-325 benchmark rather than on a disjoint held-out calibration subset. We therefore treat $T=5$ as a benchmark-tuned deployment setting rather than as a strictly held-out calibration protocol. In practice the sweep is broad rather than brittle: improvement from the uncalibrated setting to the useful regime is large and monotonic, and values in the range $T\in[3,5]$ already recover most of the gain. Nonetheless, a larger external benchmark would be preferable for fully held-out calibration selection.

\section{Prior-Only Distribution-Shift Baselines}
\label{sec:prior_baselines}

To quantify what can be achieved by label frequencies alone, we constructed prior-only baselines transferring ICSD extinction-group frequencies onto RRUFF. These baselines never inspect the diffraction pattern; they simply rank labels by their training-set frequency.

An ICSD-frequency baseline reaches approximately 9.7\% Top-1 and 27.2\% Top-5 on RRUFF-like extinction labels. Our calibrated stage-2 champion reaches 9.54\% Top-1 and 43.08\% Top-5 on the harsh real RRUFF-325 benchmark. The near-match in Top-1 shows that real-data ambiguity and class imbalance set a severe ceiling for the single best guess, but the large Top-5 gap ($+$15.9 percentage points) shows that the model is not behaving like a static histogram: it is using diffraction evidence to re-rank candidates on a per-pattern basis.

These prior-only baselines also help interpret comparisons with earlier literature. At low $k$, trained neural networks remain comfortably above the ICSD-frequency prior, indicating genuine feature learning. However, at higher $k$ the blind ICSD prior can match or exceed the ranking of uncalibrated models trained on ICSD-like distributions, suggesting that the relevant issue is not simply whether a model beats a frequency baseline at Top-1 but whether its lower-ranked predictions remain calibrated under distribution shift from synthetic to real data. Our calibration study addresses this directly: temperature scaling restores usable uncertainty so that the external empirical prior improves ranking breadth rather than merely reinforcing an already overconfident internal prior.

\section{Corrected RRUFF-Conditioned Generator}
\label{sec:generator}

The final RRUFF-conditioned stage-2 datasets used in our largest curriculum runs were generated with a corrected crystal-structure pipeline rather than with unconstrained chemistry replay alone. The main-phase stoichiometry is first constructed from Wyckoff-compatible multiplicities for the target space group, so that the per-species count vector is symmetry-compatible by construction before PyXtal placement begins. This removes the dominant early rejection mode seen in our first realistic-generator attempts, where many compositions were mathematically incompatible with the target Wyckoff multiplicities.

We further apply three constraints necessary for reliable label integrity:
\begin{enumerate}
\item Unsupported or pathological elements are removed from the chemistry support, including elements for which the PyXtal radius tables are incomplete.
\item For the main phase, at least one general-position orbit is required when available, reducing accidental symmetry elevation into higher-symmetry supergroups.
\item The generated main phase is accepted only if the realized extinction group detected by \texttt{spglib} matches the intended label.
\end{enumerate}

For the impurity phase, exact symmetry preservation is not required; the impurity only needs to yield a physically plausible additional powder pattern. The final stage-2 production recipe also used a tighter \texttt{spglib} tolerance ($\texttt{symprec}=10^{-3}$ rather than $10^{-2}$), which substantially reduced false symmetry elevation during validation while improving throughput.

\begin{table}[H]
\caption{Nuisance factors in the corrected RRUFF-conditioned generator. The generator is realism-oriented but not exhaustive: it models line-shape variation, backgrounds, impurity structure, and stochastic amplitude distortion, while leaving several instrument-specific effects for future work.}
\begin{ruledtabular}
\begin{tabular}{lll}
\parbox[t]{0.28\linewidth}{Component} & \parbox[t]{0.30\linewidth}{Modeled in the current generator} & \parbox[t]{0.30\linewidth}{Not yet modeled explicitly} \\
\colrule
\parbox[t]{0.28\linewidth}{Peak profile / resolution} & \parbox[t]{0.30\linewidth}{Empirical Caglioti-like broadening coefficients $(U,V,W,X,Y,\eta)$ sampled from family-conditioned priors} & \parbox[t]{0.30\linewidth}{Instrument-to-instrument geometry changes beyond the fixed Cu-K$\alpha$ lab setup} \\
\parbox[t]{0.28\linewidth}{Background} & \parbox[t]{0.30\linewidth}{Chebyshev polynomial background with coefficient jitter; additional smooth Gaussian-process baseline warp} & \parbox[t]{0.30\linewidth}{Fluorescence-specific backgrounds; substrate-specific scattering models} \\
\parbox[t]{0.28\linewidth}{Noise} & \parbox[t]{0.30\linewidth}{Additive Gaussian noise plus multiplicative intensity noise} & \parbox[t]{0.30\linewidth}{Detector-specific count statistics beyond this approximate noise family} \\
\parbox[t]{0.28\linewidth}{Intensity distortions} & \parbox[t]{0.30\linewidth}{Smooth amplitude distortion field to mimic family-dependent relative-intensity warping} & \parbox[t]{0.30\linewidth}{Explicit preferred-orientation models with crystal-habit parameters} \\
\parbox[t]{0.28\linewidth}{Peak positions} & \parbox[t]{0.30\linewidth}{Fixed Cu-K$\alpha$ wavelength and fixed 8501-point $2\theta$ grid; zero offset held at $0$ in the production recipe} & \parbox[t]{0.30\linewidth}{Sample displacement / zero-shift variability across instruments; wavelength-conditioned inference} \\
\parbox[t]{0.28\linewidth}{Impurities / mixtures} & \parbox[t]{0.30\linewidth}{Additional physically plausible impurity phase patterns mixed into the main phase, with 70\% of patterns left single-phase, 20\% assigned a weak impurity fraction of 0.5--3\%, and 10\% assigned a stronger impurity fraction of 3--10\%} & \parbox[t]{0.30\linewidth}{Fully instrument-conditioned multi-phase refinement effects} \\
\end{tabular}
\end{ruledtabular}
\label{tab:generator_nuisance}
\end{table}

\section{Training Lineage and Final Stage-2 Runs}
\label{sec:lineage}

\begin{table}[H]
\caption{Checkpoint labels used in the main text and their corresponding Weights \& Biases run identifiers.}
\begin{ruledtabular}
\begin{tabular}{ll}
Main-text label & W\&B / checkpoint ID \\
\colrule
Stage-1 (uniform only) & \texttt{ic6gfmvm} \\
Stage-2a (4.2M + fine-tune) & \texttt{h7dqlpfp} \\
Stage-2b (larger uniform + fine-tune) & \texttt{guvsasrs} \\
Legacy (replay-era) & \texttt{wwuvp1kj} \\
350k baseline & \texttt{7pv3pv3y} \\
Stage-2c (corrected fine-tune) & \texttt{9rwv1qly} \\
Stage-2c-aux025 (ablation) & \texttt{0879pvmm} \\
Fig.~S5 ViT Attention & \texttt{pi7r8pah} \\
\colrule
Table~I SPG ResNet-18$^\star$ & \texttt{ve988dop} \\
Table~I / Fig.~S1 Ext ResNet-18 (990k)$^\star$ & \texttt{t75aaxpz} \\
Fig.~S1 Ext 99k$^\star$ & \texttt{d7bovz1p} \\
Fig.~S1 Ext 396k$^\star$ & \texttt{9ytnffm7} \\
Fig.~S1 Ext 792k$^\star$ & \texttt{q3hxsqw8} \\
Fig.~S1 Ext 1.48M$^\star$ & \texttt{wj5yemvc} \\
Fig.~S1 Ext 2.17M$^\star$ & \texttt{fmwj0hcj} \\
Fig.~S1 Ext 3.16M$^\star$ & \texttt{ujr9w21e} \\
Fig.~S1 Ext 3.76M$^\star$ & \texttt{eiyfd7gf} \\
Fig.~S1 Ext 4.15M$^\star$ & \texttt{srq8yg04} \\
Fig.~S1 Ext 5.14M$^\star$ & \texttt{jek05aky} \\
Table~S2 / S8 Balanced$^\star$ & \texttt{ivvi2z1j} \\
Table~S2 / S8 ICSD Bias$^\star$ & \texttt{m9mdi5p9} \\
Table~S8 Augmented$^\star$ & \texttt{kiifq20z} \\
Table~S9 Non-DeDup$^\star$ & \texttt{u9quo4zn} \\
Table~S9 DeDup$^\star$ & \texttt{xhx3ex54} \\
Fig.~S2$^\star$ Top Ext. Group Performance & \texttt{rbwbgj89} \\
Fig.~S6$^{\star\star}$ CNN Interp & \texttt{kiifq20z}
\end{tabular}
\end{ruledtabular}
\vspace{2pt}
{\footnotesize $^\star$\,CNN checkpoint files are named \texttt{xrd\_model\_\{ID\}.pth}. $^{\star\star}$\,Model used to produce graphs, source code on GitHub}
\label{tab:checkpoint_mapping}
\end{table}
The final curriculum evaluated in this work used the following lineage:
\begin{enumerate}
\item \textbf{Uniform stage-1 pretraining}: a merged uniform synthetic dataset of 1,381,805 samples, yielding the Stage-1 checkpoint (\texttt{ic6gfmvm}).
\item \textbf{Corrected RRUFF-conditioned stage-2 fine-tuning}: a merged train-ready dataset of 2,346,400 samples assembled from one local corrected 100k run and three external shard sets (249.6k, 998.4k, and 998.4k).
\item \textbf{Low-learning-rate domain adaptation}: the Stage-1 checkpoint was fine-tuned for 3 epochs on the corrected 2.346M stage-2 dataset at learning rate $10^{-5}$, producing the Stage-2c checkpoint (\texttt{9rwv1qly}).
\item \textbf{Aux-loss ablation}: a matched run with reduced auxiliary extinction-group loss weight (\texttt{aux025}) produced the Stage-2c-aux025 checkpoint (\texttt{0879pvmm}).
\end{enumerate}

On the synthetic held-out split of the merged 2.346M stage-2 dataset, Stage-2c (\texttt{9rwv1qly}) reached 13.40\% / 24.32\% / 33.61\% Top-1/3/5 on the decoded head and 33.04\% / 60.12\% / 73.46\% on the auxiliary extinction-group head. The Stage-2c-aux025 ablation slightly improved the synthetic decoded Top-1/5 to 13.71\% / 34.07\%, but did not outperform the final calibrated real-benchmark operating point.

After the main Stage-2c study was complete, we also ran an exploratory continuation from a larger uniform checkpoint. This later Stage-1-style checkpoint (\texttt{xgy3h6uw}) was trained on a 2,000,000-sample uniform synthetic dataset, then fine-tuned for 3 epochs on the same corrected 2,346,400-sample RRUFF-conditioned stage-2 dataset, producing checkpoint \texttt{2vegisf1}. On the stage-2 synthetic held-out split, \texttt{2vegisf1} reached 14.71\% / 26.34\% / 35.60\% Top-1/3/5 on the decoded head and 34.53\% / 61.61\% / 74.89\% on the auxiliary extinction-group head.

\section{Temperature Scaling and Calibration Details}
\label{sec:temperature}

The late-stage real-benchmark experiments revealed that the dominant residual failure mode of the Stage-2c model's auxiliary head was not simple forgetting of crystallographic structure, but overconfidence. Because the stage-2 model is trained on millions of RRUFF-conditioned patterns, part of the geological prior is implicitly absorbed into the weights. Adding the external empirical prior at inference without calibration therefore risks double-counting the same bias.

We evaluated post-hoc temperature scaling~\cite{guo2017calibration} on RRUFF-325. The strongest result came from the Stage-2c auxiliary head with temperature $T=5$, which improved the auxiliary path from 2.15\% Top-1 and 14.46\% Top-5 in its uncalibrated form to 9.54\% / 27.38\% / 43.08\% Top-1/3/5 without changing any model weights. This calibrated auxiliary head is the final real-benchmark champion.

\begin{table}[H]
\caption{Calibration metrics for the Stage-2c auxiliary head on RRUFF-325 before and after temperature-scaled Bayesian decoding. Confidence intervals are nonparametric bootstrap 95\% intervals over the 325-pattern benchmark.}
\begin{ruledtabular}
\begin{tabular}{lccccc}
Operating point & Top-1 (\%) & Top-5 (\%) & ECE & NLL & Brier \\
\colrule
Uncalibrated aux & 2.15 [0.62, 3.69] & 14.46 [10.77, 18.46] & 0.457 & 6.58 & 1.283 \\
Calibrated Bayesian aux ($T=5$) & 9.54 [6.46, 12.92] & 43.08 [37.85, 48.62] & 0.049 & 3.70 & 0.958 \\
\end{tabular}
\end{ruledtabular}
\label{tab:calibration_metrics_325}
\end{table}

Here ECE denotes the expected calibration error, a bin-based summary of confidence--accuracy mismatch; NLL denotes the negative log-likelihood, which penalizes overconfident wrong predictions; and the multiclass Brier score is the mean squared probability error~\cite{brier1950verification}. For a physics audience, the key point is that these metrics quantify not just whether the model is right, but whether its reported probabilities are trustworthy enough to combine with an external geological prior.

For completeness, we also explored softening the split-head logits before fusion in earlier checkpoints. Sweeping split temperature $T \in \{1.5, 2.0, 2.5, 3.0\}$ and fusion weight $\alpha \in [0.15, 0.35]$ on the Stage-2a checkpoint:

\begin{table}[H]
\caption{Temperature scaling on split-head fusion. Temperature scaling did not improve the untuned fusion Top-1 but shows promise for Top-$k$ breadth calibration.}
\begin{ruledtabular}
\begin{tabular}{ccc}
Configuration & Parameters & Accuracy \\
\colrule
Best swept Top-1 & $T = 1.5$, $\alpha = 0.35$ & 15.43\% \\
Best swept Top-5 & $T = 2.5$, $\alpha = 0.15$ & 48.84\% \\
\colrule
Untuned baseline & $T = 1.0$, $\alpha = 0.25$ & 15.64\% \\
\end{tabular}
\end{ruledtabular}
\label{tab:temperature_split}
\end{table}

The later Stage-2c-aux025 ablation was designed as a pre-hoc analogue of this calibration idea: reduce the auxiliary head's dominance during stage-2 training. On RRUFF-325, however, the best calibrated aux025 operating point still occurred at $T=5$ and reached 6.77\% / 27.08\% / 43.69\% Top-1/3/5. The ablation slightly improved Top-5 breadth but did not surpass calibrated Stage-2c on Top-1, confirming that strong auxiliary gradients are necessary during training and that calibration is best handled post-hoc.

For completeness, we also evaluated the exploratory \texttt{2vegisf1} continuation on RRUFF-325. Its best exact Top-1 operating point was the calibrated auxiliary decoder at $T=3$, reaching 7.38\% / 25.85\% / 38.15\% Top-1/3/5. Its broadest ranking occurred at $T=5$, reaching 5.54\% / 29.85\% / 46.15\% Top-1/3/5. Relative to the stronger uniform-only precursor \texttt{xgy3h6uw}, this continuation improved breadth substantially but softened the single best guess; relative to the main-text Stage-2c champion, it slightly improved Top-5 breadth while remaining weaker on Top-1. We therefore treat it as an informative late continuation rather than as a replacement for the calibrated Stage-2c deployment checkpoint.

We also tested whether this sharpness-versus-breadth tradeoff could be partially repaired without retraining by weight-space interpolation (WiSE-FT)~\cite{wortsman2022wiseft} between the uniform checkpoint \texttt{xgy3h6uw} and the later fine-tuned continuation \texttt{2vegisf1}. In practical terms, this means forming a new checkpoint by linearly interpolating the two models' learned weights rather than fusing their output probabilities. Table~\ref{tab:wiseft} shows that this simple interpolation substantially improves the Pareto frontier. In particular, $\alpha=0.7$ recovers most of the uniform model's sharper Top-1 behavior while preserving nearly all of the fine-tuned model's broad Top-5 ranking, and $\alpha=0.5$ yields the strongest overall Top-5 breadth in this late family of runs. This suggests that part of the final tradeoff comes from over-adaptation during fine-tuning rather than from an unavoidable conflict between exact Top-1 and broad calibrated ranking.

\begin{table}[H]
\caption{Late WiSE-FT interpolation between the 2.0M uniform checkpoint \texttt{xgy3h6uw} and its 2.346M RRUFF-conditioned continuation \texttt{2vegisf1}, evaluated on RRUFF-325 using calibrated auxiliary decoding. Weight-space interpolation recovers much of the uniform model's Top-1 sharpness while preserving the fine-tuned model's Top-5 breadth.}
\begin{ruledtabular}
\begin{tabular}{lcc}
Checkpoint / operating point & Top-1/3/5 (\%) & Interpretation \\
\colrule
\texttt{xgy3h6uw}, best Top-1 & 8.92 / 11.38 / 16.62 & Sharpest uniform-only operating point \\
\texttt{2vegisf1}, best Top-1 & 7.38 / 25.85 / 38.15 & Fine-tuned continuation, sharper setting \\
\texttt{2vegisf1}, best Top-5 & 5.54 / 29.85 / 46.15 & Fine-tuned continuation, broadest setting \\
WiSE-FT $\alpha=0.7$, $T=3$ & 8.62 / 24.92 / 38.77 & Best Top-1 among interpolants \\
WiSE-FT $\alpha=0.7$, $T=5$ & 7.08 / 30.46 / 46.46 & Best balanced late operating point \\
WiSE-FT $\alpha=0.5$, $T=5$ & 7.08 / 30.46 / 46.77 & Broadest Top-5 among interpolants \\
\end{tabular}
\end{ruledtabular}
\label{tab:wiseft}
\end{table}

\subsection{Matched Space-Group Control}

The reviewer-facing control for the extinction-group formulation is a matched space-group-trained model evaluated under the same data regime. We therefore relabeled the same 2.0M uniform synthetic corpus at the 230-space-group level and trained the same backbone on a clean 1.6M/200k/200k train/validation/test split. A first 1-epoch preview reached 7.08\% / 15.91\% / 22.05\% Top-1/3/5 on the held-out 230-class space-group test set, and a second effective epoch improved this to 9.32\% / 19.15\% / 25.57\% during training-time SG evaluation. We then completed the actual post-hoc SG$\rightarrow$EG collapse requested by the reviewer: for every held-out test pattern, we took the full 230-way SG probability vector, summed the probabilities of all space groups belonging to the same extinction-group equivalence class, and rescored the resulting 99-way distribution against the matched extinction-group test labels. The SG and EG held-out test sets were verified to be row-aligned ($16/16$ spot-check rows matched exactly, max absolute intensity difference $0.0$).

This final matched control strongly favors direct extinction-group training. The raw SG model reaches 7.08\% / 15.91\% / 22.05\% Top-1/3/5 in SG space, and the post-hoc SG$\rightarrow$EG collapse improves this to 8.61\% / 19.37\% / 26.76\% in EG space. However, the matched direct-EG auxiliary model on the same 2.0M held-out regime reaches 19.32\% / 34.28\% / 43.38\%. Thus, although the earlier synthetic SG-versus-EG comparison in the main text was indeed confounded by class count and dataset size, the cleaner matched control leads to the same qualitative conclusion: direct training on extinction groups is substantially better than training on 230 space groups and collapsing only at evaluation time.

\begin{table}[H]
\caption{Matched SG$\rightarrow$EG control on the same 2.0M uniform corpus and held-out split. Collapsing the 230-way SG probabilities into extinction-group space improves over raw SG scoring, but remains far below direct extinction-group training on the same regime.}
\begin{ruledtabular}
\begin{tabular}{lccc}
Matched control / evaluation space & Top-1 (\%) & Top-3 (\%) & Top-5 (\%) \\
\colrule
SG-230 model, raw SG space (1 epoch checkpoint) & 7.08 & 15.91 & 22.05 \\
SG-230 model, post-hoc collapsed to EG space & 8.61 & 19.37 & 26.76 \\
Direct EG model, matched 99-way auxiliary head & 19.32 & 34.28 & 43.38 \\
\end{tabular}
\end{ruledtabular}
\label{tab:sg230_control}
\end{table}

\section{Classical Pawley Baseline}
\label{sec:pawley}

We developed a classical statistical baseline for extinction-group inference using sparse Pawley fitting. The original full Bravais-from-scratch sweep proved too unstable for practical use (typically 3--5 hours per pattern, with frequent convergence to incorrect lower-symmetry solutions). We therefore reformulated the problem as a \emph{topology-guided conditional benchmark}: the calibrated neural model provides a local subgroup-DAG neighborhood, and the classical backend ranks only the extinction groups within that bounded branch. This neural prior is not only a statistical guide but also a practical accelerator: on the supported non-monoclinic branches, the bounded conditional runs complete in tens of seconds to minutes rather than requiring a multi-hour global sweep.

The first clean result is a broader supported non-monoclinic follow-up on 34 RRUFF-325 cases spanning the currently implemented $hR$ and $cF$ branches. All 34 bounded runs completed under the conditional benchmark, with zero execution errors or timeouts (Table~\ref{tab:topology_pawley_nonmp}). Exact top-ranked recovery is concentrated in the rhombohedral branch: 8/25 $hR$ cases return the exact reference space-group string, while the remaining $hR$ cases usually stay within nearby rhombohedral alternatives such as $R32$, $R\bar{3}$, or $R\bar{3}c$. The cubic face-centered branch is more ambiguous: all 9 $cF$ cases complete, but the true target is not ranked first by exact string match in any case. Even there, however, the misses remain local to the cubic face-centered branch rather than jumping to unrelated symmetries: one common return is $Fd\bar{3}$, a lower-symmetry subgroup of $Fd\bar{3}m$, while the others stay within nearby $cF$ alternatives such as $F4132$ and $Fm\bar{3}m$.

\begin{table}[H]
\caption{Topology-guided conditional Pawley follow-up on the supported non-monoclinic subset of RRUFF-325. All 34 bounded runs completed. Exact string-match recovery is strongest in the rhombohedral branch, while the cubic face-centered branch remains locally ambiguous even after bounded search.}
\begin{ruledtabular}
\begin{tabular}{lccc}
Branch & Cases & Exact top candidate & Interpretation \\
\colrule
$hR$ & 25 & 8/25 & Often exact or locally sensible within the rhombohedral branch \\
$cF$ & 9 & 0/9 & Locally ambiguous among nearby cubic face-centered alternatives \\
Total & 34 & 8/34 & No timeouts or execution errors in the bounded follow-up \\
\end{tabular}
\end{ruledtabular}
\label{tab:topology_pawley_nonmp}
\end{table}

Primitive monoclinic was the difficult branch. In the original sparse elastic-net backend, all four $mP$ pilot cases stalled before completing even the first candidate fit. Debugging showed that the bottleneck was not in the wrapper or candidate generation but in the inner sparse solve itself: primitive monoclinic patterns generate a very dense reflection basis, and the resulting profile matrix becomes numerically difficult for the original $\ell_1$-regularized solver. We therefore introduced three monoclinic-specific stabilizations for probe mode: (i) hard truncation to low angles ($2\theta \le 50^\circ$), (ii) reflection clustering before the inner solve, and (iii) a nonnegative NNLS/ridge-style solver in place of the original sparse elastic-net objective.

These changes make the monoclinic backend computationally tractable. In the initial 4-case pilot, all four $mP$ cases complete under the stabilized backend. In a broader supported follow-up over 61 monoclinic RRUFF-325 cases, 51 produced final bounded rankings under the same stabilized setting. We do \emph{not} yet treat these monoclinic rankings as a correctness result. The two Langite pilot cases (true space group $Pc$) currently rank nearby monoclinic alternatives such as $P2/c$ or $P2_1/c$ above the exact target. This is still a physically local outcome: $Pc$ is a subgroup of $P2_1/c$, so a bounded fall from $P2_1/c$ to $Pc$ lies within the expected lower-symmetry branch. By contrast, the Tintinaite cases are not appropriate correctness tests for the monoclinic backend at all. Their true space group is orthorhombic $Pnnm$, but the topology-guided pilot routed them into an $mP$ branch because the neural anchor was already on the wrong macroscopic family. For Tintinaite, the current stabilized runs therefore demonstrate computational tractability of the bounded backend, not scientific correctness of the monoclinic anchor.

\begin{table}[H]
\caption{Current status of the stabilized primitive-monoclinic follow-up. The main result at present is computational tractability: the stabilized backend completed the original 4-case pilot and 51/61 cases in a broader supported follow-up. Only the Langite cases provide a meaningful local monoclinic-ranking test.}
\begin{ruledtabular}
\begin{tabular}{llll}
Case / summary & True SG & Current top candidate & Interpretation \\
\colrule
Broader supported follow-up & 61 $mP$ cases & 51 completed rankings & Tractability established at moderate scale \\
Tintinaite R110046-9 & $Pnnm$ & $P2_1/c$ & Wrong-branch anchor; tractability only \\
Tintinaite R110163-9 & $Pnnm$ & $P2_1/c$ & Wrong-branch anchor; tractability only \\
Langite R070316-9 & $Pc$ & $P2/c$ & Local monoclinic miss \\
Langite R120042-9 & $Pc$ & $P2/c$ or $P2_1/c$ & Local monoclinic miss \\
\end{tabular}
\end{ruledtabular}
\label{tab:topology_pawley_mp_status}
\end{table}

Taken together, these experiments support a clear division of labor. The neural model is a robust proposer that localizes the crystallographic search to a small topological neighborhood. Classical Pawley verification is then practical and often useful on higher-symmetry branches, while primitive monoclinic remains the numerically hardest regime and requires family-specific stabilization. This is exactly the setting in which the learned topological prior adds value: it converts an intractable global classical search into a bounded local verification problem, while also revealing where conventional iterative optimization still struggles most severely.

\section{RRUFF Benchmark Curation and Provenance}
\label{sec:curation}

The benchmark curation pipeline is intentionally separated from the stage-2 generator. On the frozen RRUFF snapshot used for this paper, the Cu-K$\alpha$ metadata-qualified pool is first reduced to 1981 scans on the standard 8501-point grid. The large reduction from that pool to RRUFF-473 is therefore not a trivial metadata screen but the family-coherence curation step itself. Real RRUFF scans are grouped by mineral family, split by mineral-plus-space-group assignment, and then clustered by tight lattice compatibility. Singleton clusters are admitted only when they remain similar to the main family cluster in both lattice parameters and pattern shape, while the small residual set of partial clusters is resolved primarily by within-cluster medoid coherence, with similarity to the main family cluster used as a secondary tie-break. Families that do not support a stable coherent core are excluded entirely, while retained families may keep more than one scan when repeated measurements remain internally consistent rather than trivially duplicate. The final RRUFF-473 benchmark therefore keeps representative/core scans from tight and mixed families while excluding strongly heterogeneous families.

The nuisance-fit labels (\emph{recoverable}, \emph{usable}, \emph{poor}, \emph{catastrophic}) come from downstream profile-fitting diagnostics rather than from human judgment. These labels were used only for stratified analysis, not as training targets. The broader RRUFF-473 benchmark intentionally preserves difficult cases rather than removing them, because the goal is to establish an honest sim-to-real baseline rather than an artificially sanitized leaderboard. The downstream RRUFF-325 benchmark is then a deterministic subset of RRUFF-473 obtained by retaining only the usable-or-better and recoverable strata under frozen $R_{\mathrm{wp}}$ thresholds. This separation is deliberate: RRUFF-473 is the realism benchmark, while RRUFF-325 is the cleaner operating slice used for calibration and topology analyses.

\section{Database Architecture}
\label{sec:database}

The database for this project uses a relational SQLite index for structural metadata (space group, lattice parameters, atomic positions, extinction information, provenance) with diffraction patterns stored separately in HDF5 files. To support parallel generation across compute nodes, the pipeline writes to independent SQLite databases that are merged post-generation using deterministic SQL operations. An atomic-write pattern (writing to a \texttt{.partial.<pid>} sibling file, then renaming) prevents corruption from interrupted HDF5 merges.

\section{Positional Encoding Ablation on the Mixed-200k Real-Data Pilot}
\label{sec:positional_ablation}

We ran a matched $2\times2$ positional ablation on the mixed-200k pilot checkpoint family, comparing the full model (coordinate channel + physics-aware positional encoding) against three bounded continuations: physics-aware positional embedding only, coordinate channel only, and no explicit positional mechanism. All variants were evaluated with the same calibrated Bayesian auxiliary decoder on RRUFF-325 and RRUFF-473, and split-head validity was measured separately as the fraction of patterns whose decoded rule bits mapped to exactly one legal extinction-group template.

\begin{table}[H]
\caption{Mixed-200k positional ablation on the real-data benchmarks. The auxiliary calibrated head remains surprisingly competitive even when positional mechanisms are removed, but strict split-head validity collapses completely without the additive physics-aware positional encoding. This indicates that the coordinate channel alone is not sufficient to preserve legal rule decoding, while the physics-aware positional embedding carries the more important positional prior in the structured decoder.}
\begin{ruledtabular}
\begin{tabular}{llcccc}
Variant & Benchmark & Top-1 (\%) & Top-5 (\%) & ECE (\%) & Split valid (\%) \\
\colrule
Full model & RRUFF-325 & 13.23 & 40.92 & 1.31 & 1.54 \\
Physics-aware PE only & RRUFF-325 & 13.85 & 39.38 & 3.34 & 1.23 \\
Coordinate channel only & RRUFF-325 & 12.62 & 41.85 & 4.29 & 0.00 \\
No explicit positional mechanism & RRUFF-325 & 12.62 & 43.38 & 4.97 & 0.00 \\
\colrule
Full model & RRUFF-473 & 13.53 & 49.05 & 0.95 & 1.27 \\
Physics-aware PE only & RRUFF-473 & 13.95 & 47.57 & 1.85 & 1.48 \\
Coordinate channel only & RRUFF-473 & 12.68 & 48.41 & 3.01 & 0.00 \\
No explicit positional mechanism & RRUFF-473 & 12.90 & 51.16 & 4.54 & 0.00 \\
\end{tabular}
\end{ruledtabular}
\label{tab:positional_ablation}
\end{table}

These results sharpen the architectural interpretation in the main text. The coordinate channel clearly remains useful as an input-side physical ruler, but it is not the mechanism that preserves rule-consistent structured decoding under real-data transfer. The additive physics-aware positional embedding is the more important positional prior for the split path: when it is removed, legal split-head decoding disappears entirely, even though the auxiliary calibrated ranking remains nontrivial. Conversely, retaining the physics-aware positional embedding while dropping the coordinate channel preserves most of the weak but nonzero split-head validity of the full mixed-200k pilot, albeit with degraded calibration.

\begin{figure}[H]
\centering
\includegraphics[width=0.82\textwidth]{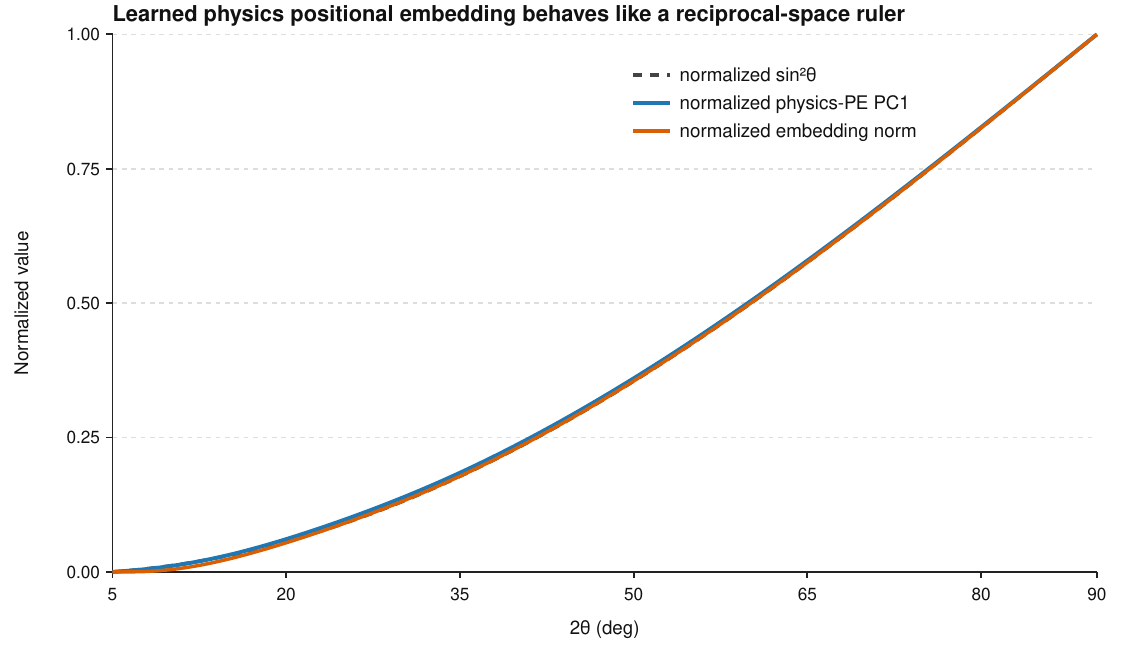}
\caption{Learned physics-aware positional embedding in the final real-data model. The plot compares the normalized input $\sin^2\!\theta$ coordinate to the first principal component of the learned patch-level physics embedding and to the embedding norm. The learned physics term is almost entirely one-dimensional (PC1 explains 99.98\% of the embedding variance) and is monotone in the input $\sin^2\!\theta$ coordinate, indicating that the MLP has learned a smooth reciprocal-space ruler rather than an arbitrary high-dimensional positional code. Expressed in reciprocal-space units, the leading component is almost perfectly aligned with $Q^2$ ($r = 0.99997$).}
\label{fig:physics_pe_ruler}
\end{figure}

This diagnostic clarifies the role of the additive physics-aware positional term. Although the embedding itself lives in the full token dimension, the trained mapping collapses almost entirely onto a single monotone direction aligned with the underlying diffraction coordinate. In other words, the physics-aware positional MLP is not inventing a complicated token identity code; it is learning a smooth scalar warping of reciprocal-space position, effectively a $Q^2$-like ruler, that can then be combined with the learned absolute embedding used by the transformer backbone. This sharpens the ablation interpretation in Table~S3: the benefit of the physics-aware positional term is not extra embedding complexity, but the fact that it supplies the transformer with the correct reciprocal-space ruler at the patch-token level.

\bibliography{references_revised}

\end{document}